\documentclass[aip,twocolumn,amsmath,amssymb,reprint]{revtex4-1}
\usepackage{graphicx}
\usepackage{bm}
\usepackage[capitalise]{cleveref} 
\usepackage[labelfont=bf, font={normal}, justification=justified]{caption} 
\usepackage[labelfont=rm, font={small}, justification=justified]{subcaption} 
\usepackage{ragged2e} 
\DeclareCaptionJustification{justified}{\justifying} 
\usepackage[locale = UK,product-units = brackets,
			list-pair-separator = {;},detect-all]{siunitx} 
\protected\def\dmtr{\ensuremath{\varnothing}} 
\DeclareSIUnit{\dimless}{\ensuremath{\text{---}}} 
\DeclareSIUnit{\pixel}{\ensuremath{\text{px}}} 
\usepackage{fontawesome} 
\usepackage{mathtools} 
\usepackage{dutchcal} 
\usepackage{chemformula} 
\usepackage{upgreek} 
\usepackage{derivative} 
\usepackage{multirow} 
\usepackage{accents} 
\usepackage{stmaryrd} 
\usepackage{mathrsfs}

\usepackage{booktabs}  
\usepackage{amsmath}    
\usepackage{array}      
\usepackage[bb=fourier]{mathalpha} 
\let\fourierbb\mathbb
\AtBeginDocument{%
    \let\mathbb\relax
    \newcommand{\mathbb}[1]{\fourierbb{#1}} 
}
\newcommand{\vek}[1]{\ensuremath{\mathbf{#1}}} 
\newcommand{\mat}[1]{\ensuremath{\mathbf{#1}}} 
\newcommand{\upe}{\ensuremath{\mathrm{e}}} 
\newcommand{\upi}{\ensuremath{\mathrm{i}}} 
\newcommand{\note}[1]{\ensuremath{\texttt{\tiny{#1}}}} 

\newcommand{\braket}[2]{\ensuremath{\big\langle{#1}\big|{#2}\big\rangle}}

\newcommand{\braketo}[3]{\ensuremath{\big\langle{#1}\big|{#2}\big|{#3}\big\rangle}}

\begin{document}
\preprint{AIP/123-QED}
\title[Extraction of Bend-Resolved Modal Basis in Deformed Multimode Fiber]{Extraction of Bend-Resolved Modal Basis in Deformed Multimode Fiber}
\author{L.~Skvarenina}
\email[\:\faEnvelopeO\:]{\href{mailto:Lubomir.Skvarenina@glasgow.ac.uk}{Lubomir.Skvarenina@glasgow.ac.uk}}
\affiliation{Structured Photonics, James Watt School of Engineering, University of Glasgow, 11 Chapel Lane, Glasgow, G11\,6EW, United Kingdom}
\affiliation{Department of Physics, Faculty of Electrical Engineering and Communication, Brno University of Technology, Technick\'{a}~2848/8, Brno, 616\,00, Czech Republic}
\author{S.~Simpson}
\affiliation{Microphotonics, Institute of Scientific Instruments of the Czech Academy of Sciences, Kr\'{a}lovopolsk\'{a}~147, Brno, 612\,00, Czech Republic}
\author{Y.~Alizadeh}
\affiliation{Structured Photonics, James Watt School of Engineering, University of Glasgow, 11 Chapel Lane, Glasgow, G11\,6EW, United Kingdom}
\author{M.~P.~J.~Lavery}
\affiliation{Structured Photonics, James Watt School of Engineering, University of Glasgow, 11 Chapel Lane, Glasgow, G11\,6EW, United Kingdom}
\date{12\,June\,2025}
\begin{abstract}%
Mode mixing in optical fibers caused by mechanical bending induces perturbations that distort the spatial field profile of coherent beams as they propagate through few-mode or multimode fibers. The observed output from a~bent fiber commonly appears as complex speckle, which is challenging to relate directly to the underlying deformation, particularly in continuously varying systems such as aerially deployed fibers or fiber-integrated sensors in mechanical structures. We introduce a~novel method for constructing a~complete deformation-resolved orthonormal modal basis that captures the optical response of a~multimode fiber across a~range of controlled mechanical deformations. The basis is derived via a~two-stage singular value decomposition framework that initially constructs deformation-specific orthonormal mode sets from speckle pattern correlation matrices and  subsequently decomposes the aggregated sets to produce a~unified functional basis that comprehensively spans the deformation-induced modal subspace supported by the fiber. This hierarchical framework yields an energy-balanced representation that isolates statistically dominant field components across all deformation states, approximates superpositions of the fiber’s propagation-invariant modes, systematically encodes deformation-induced perturbations, and supports robust decomposition of output fields across varying mechanical conditions. Such a~basis enables tracking of mechanically induced modal evolution in deployed fibers, supporting distributed sensing, network resilience, and predictive fault diagnostics, with potential for integration into mode-division multiplexing systems.
\end{abstract}
\maketitle 

\section{\label{sec:introduction}Introduction}%

\noindent In-network fiber sensing has emerged as a~rapidly advancing field in optical research and technology, with applications in structural health monitoring, robotic sensing, environmental observation, earthquake early warning, and prediction of network outages~\cite{Lu:2019,Galloway:2019,Hussain:2023,Lior:2023,Zhang:2023}. As global data capacity demands continue to grow, multi-mode fiber (MMF) communication technologies are being actively developed to surpass the capacity and energy efficiency limitations of conventional single-mode fibers (SMFs)~\cite{Essiambre:2012,Tkach:2010,Nakazawa:2022}. This progress presents an opportunity to develop new sensing technologies that leverage the unique optical interactions in MMFs, or even few-mode fibers (FMFs), for detecting mechanical deformations along the fiber. Unlike SMFs, MMFs exhibit mode-dependent loss (MDL), group delay (GD) spread due to modal dispersion, and intermodal coupling, all of which contribute to signal distortion and errors in multiplexed data transmission~\cite{Winzer:2017,Juarez:2012}. Intermodal crosstalk (XT), which inherently transfers energy between adjacent spatial modes during propagation through a~shared physical channel, can further degrade or even fully disrupt transceiver-encoded information. To compensate for these impairments, multiple-input multiple-output (MIMO) digital signal processing (DSP) techniques are employed to track channel dynamics and accurately separate parallel data streams. Meanwhile, FMFs supporting a~limited number of spatial modes offer an optimal balance between achieving high data rates over long distances and reducing computational complexity, making MIMO-DSP more manageable~\cite{Klaus:2022,Sillard:2014}. Spatial super-channel architectures employ MIMO-DSP to computationally determine how power is transferred between modes in the fiber and to construct orthogonal modes between transmitter and receiver for information encoding~\cite{Hout:2024,Zhou:2021,Winzer:2018}. The modal distortions observed at the output of a~FMF or MMF are intrinsically governed by physical deformations along the fiber, particularly bending-induced variations in modal coupling, modal phase delays, and MDL. These distortions are often complex and difficult to interpret, and accurately characterizing or reconstructing the underlying relationship between mechanical perturbation and optical response remains a~fundamental challenge. Resolving this inverse problem typically requires advanced computational techniques capable of extracting meaningful structural or transmission-related information from the distorted output field~\cite{Liu:2020,Wang:2022,Bender:2023,Zheng:2024}.%


\begin{figure*}[t!]%
  	\subcaptionbox{\label{fig:volumetricSpeckle_Fiber}}{\includegraphics[width=0.73365\textwidth]{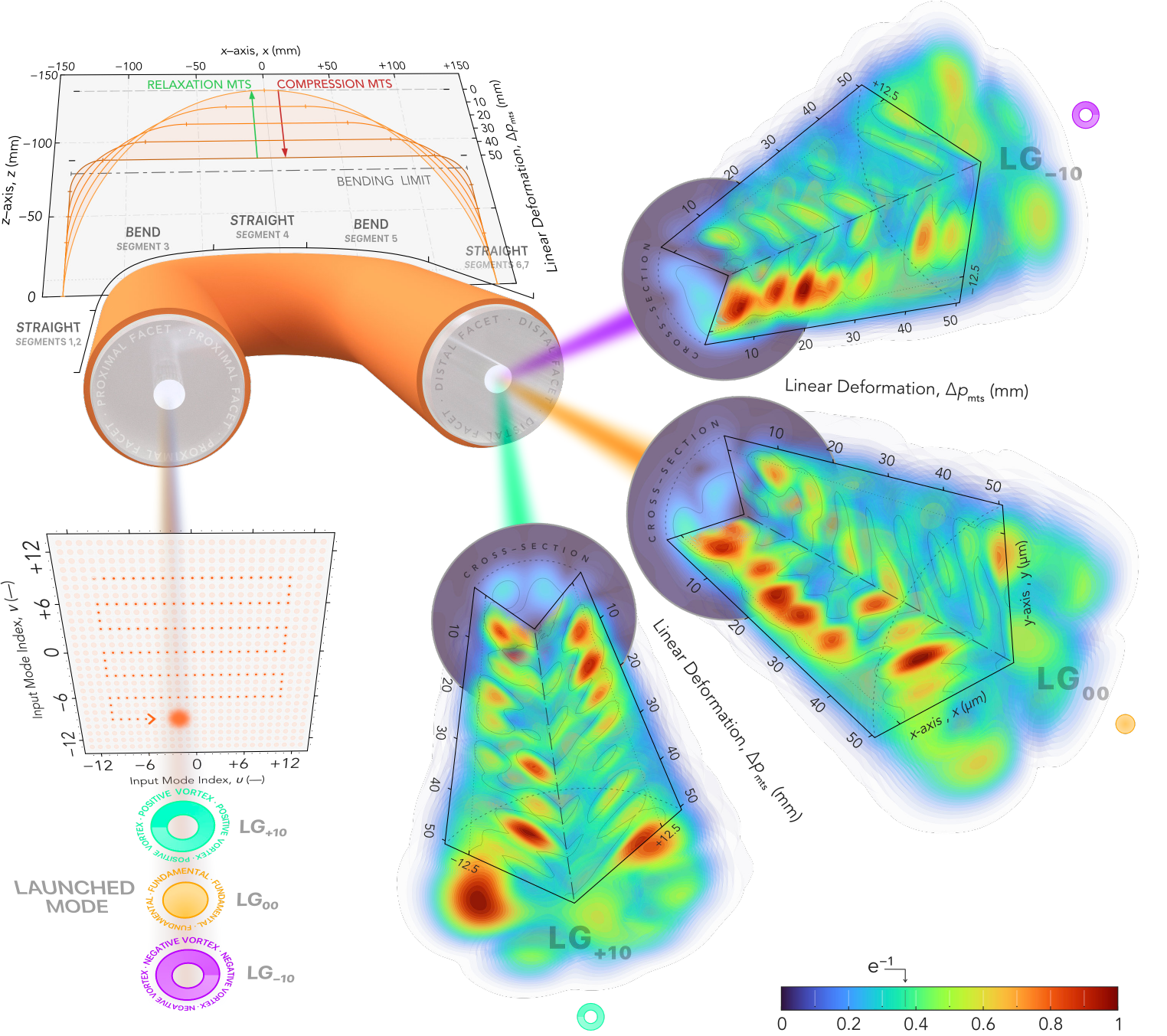}}\hfill
   	\subcaptionbox{\label{fig:powerThroughFiber_Field}}{\includegraphics[width=0.26634\textwidth]{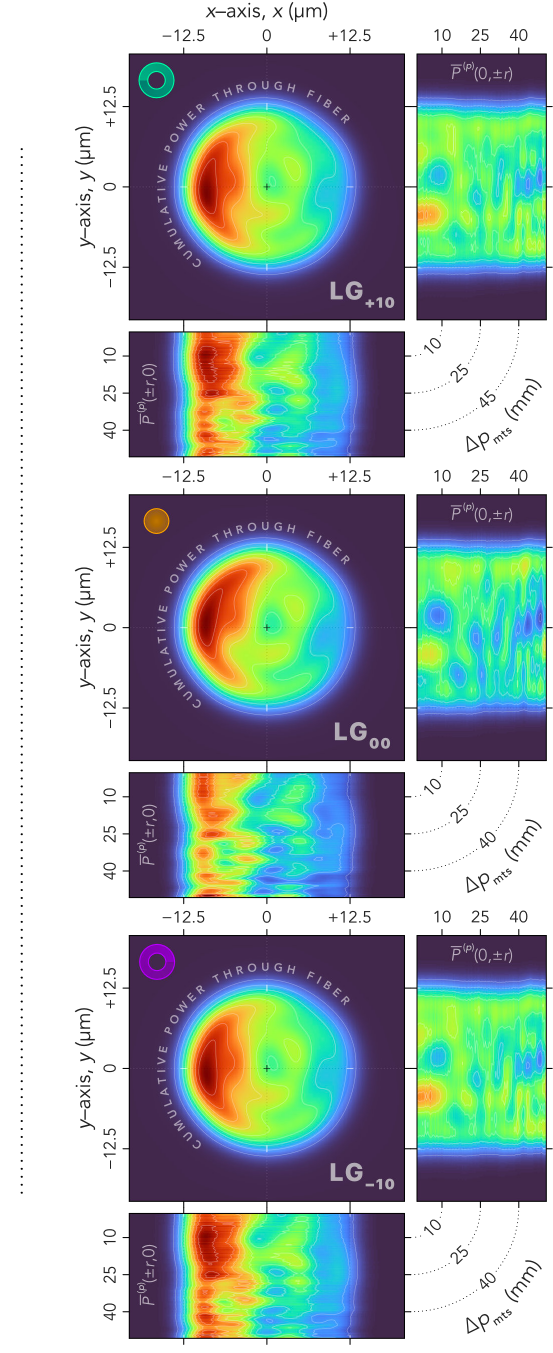}}
   	\setcounter{figure}{1}
	\caption{\label{fig:experimental_concept}Illustration of the experimentally obtained field emerging from a linearly deformed fiber under coupling conditions $\bigl\{ \mathrm{LG}_{\ell 0} \mid \ell \in \{ -1, 0, 1 \} \bigr\}$. (a)~Volumetric visualization of the evolving speckle amplitude at the distal facet, resulting from fixed proximal facet coupling at $(\upsilon,\nu) = (-3, -9)$,  for deformation states $\Delta p_{\text{mts}} = \bigl\{0.5\cdot(p-1) \mid p \in \mathbb{N},\, p \leq N_\mathrm{B}\bigr\}\,\unit{\milli\metre}$, where $p \in \left\{1, \dots, N_{\mathrm{B}}\right\}$ indexes the deformation state. Highlighted cross-section shows the amplitude quarter-volume $\bigl|S_{\mu}^{(p)}\!(x_0, y_0)\bigr|$ for $(x_0, y_0) = \bigl\{(-r, 0)\mathbin{,}\allowbreak(0, r)\bigr\}$. (b)~Cumulative power distribution averaged over all input displacements and deformations, $\overline{P}(r,\theta) = (N_\mathrm{B}N_\mathrm{S})^{-1}\allowbreak\sum_{p=1}^{N_\mathrm{B}}\allowbreak\sum_{\mu=1}^{N_\mathrm{S}}\allowbreak\bigl|{S_{\mu}^{(p)}\!(r,\theta)}\bigr|^{2}$, with corresponding cross-sections showing per-deformation contributions, $\overline{P}^{(p)}\!(x_0,y_0)\mathbin{=}\allowbreak{N}_\mathrm{S}^{-1}\sum_{\mu=1}^{N_\mathrm{S}}\big|{S_{\mu}^{(p)}\!(x_0,y_0)}\big|^{2}$ for $(x_0, y_0) \in \bigl\{(\pm{r}, 0), (0,\pm{r})\bigr\}$, where $N_{\mathrm{B}}=101$ is the number of bend configurations, and $N_{\mathrm{S}}=729$ is the number of speckle patterns per deformation.}%
\end{figure*}%


A~widely adopted method for analyzing mode propagation through complex optical systems is singular value decomposition (SVD)~\cite{Seyedinnavadeh:2024,Miller:2013}. This linear algebraic technique is typically applied to the system’s transmission matrix (TM), which describes the complex amplitude transformation of a~predefined set of input spatial modes as they propagate through the channel~\cite{Yadav:2024}. The TM is experimentally determined by injecting known input modes, selected from complete sets such as Laguerre--Gaussian (LG), Hermite--Gaussian (HG), or fiber-specific linearly polarized (LP) modes, as well as spot-like excitations depending on the optical platform, and recording their corresponding output fields. Performing SVD on the TM yields a~set of orthogonal input–output mode pairs that diagonalize the system’s response, defining a~modal basis in which energy transfer occurs through decoupled, optimally transmitting spatial channels~\cite{Kettlun:2021}. While these modes offer maximal transmission with zero crosstalk in the decoupled SVD basis, they are not physically unique and inevitably change under perturbations, as they reflect only a~specific channel realization and instance of XT at the time of measurement, and do not generally convey intrinsic physical properties of the system~\cite{Rothe:2023,Stellinga:2021}. Modeling intermodal coupling in optical fibers under varying configurations, particularly those with continuously changing bend profiles, remains a~significant challenge. Although full TM acquisition in FMF and MMF is now feasible through high-speed, phase-resolved techniques~\cite{Stiburek:2023,Jakl:2022}, reconciling these experimental measurements with numerical models typically requires discretizing the fiber into segments with near-constant curvature and variable lengths to capture localized geometric variations and preserve accuracy in modeling intermodal coupling~\cite{Resisi:2020,Li:2021:a}. Due to the non-commutative nature of TMs, each fiber segment must be modeled independently, and the cumulative modal evolution is accurately captured only through sequential matrix multiplication, as direct averaging fails to preserve the order-dependent transformations essential to mode propagation~\cite{Matthes:2021,Cao:2023}. This complexity is further compounded by micro\-bending and stress-induced birefringence, which necessitate high spatial resolution and detailed refractive index profiling to resolve localized polarization effects and variations in modal coupling. Accurately modeling these phenomena requires finely discretized, spatially varying TMs that account for the resulting micro-scale anisotropies and index perturbations. While such impairments are effectively managed in MIMO-DSP-based coherent communication systems through dynamic TM estimation and real-time equalization, these techniques do not recover or interpret the underlying physical perturbations along the fiber~\cite{Ryf:2012,Zhang:2025}. This limitation motivates the development of complementary, physically interpretable methods that operate directly on output field statistics to reconstruct deformation-resolved modal structure and enable integrated sensing alongside high-capacity transmission~\cite{Rothe:2020,Amitonova:2020}.%


In this paper, we introduce a~systematic framework for constructing a~stable, bend-resolved modal basis by applying a~two-stage SVD to correlation matrices computed from speckle field statistics generated by a~multimode fiber subjected to systematically controlled bending conditions. A~set of spatially displaced, diffraction-limited foci incorporating orbital angular momentum (OAM)-carrying beams is employed to evaluate the influence of input excitation variations on the intermediate singular modes and to demonstrate the robustness of the resulting bend-resolved basis to differences in coupling conditions. Each mechanical deformation is treated as a~structured optical transformation, with its corresponding field statistics contributing to a~localized orthonormal basis. A~global SVD then compacts and reorients these bases into a~single, stable coordinate system that spans the dominant subspace of output field variations across all sampled deformations. This empirically derived basis captures both the intrinsic modal content of the fiber and the perturbations induced by bending, without requiring knowledge of the wave equation, propagation constants, or refractive index profile. Unlike conventional singular modes defined through TM at fixed configurations, which are orthogonal but not physically invariant, our bend-resolved modes are consistently aligned with the physical deformation of the system and reflect coherent structural evolution across different bending states~\cite{Mididoddi:2025}. Moreover, they provide physically interpretable signatures of mechanical perturbations and enable early detection of fiber strain, interference, or damage. The method is directly compatible with coherent mode-division multiplexing (MDM) systems, where real-time TM estimation is already performed for known spatial inputs. Integrated with machine learning and field calibration, it enables transceiver-level sensing functionality, enhancing physical-layer resilience in communication networks while supporting structural health monitoring in fiber-deployed infrastructures such as bridges, tunnels, and rail corridors~\cite{Kishida:2025,Zhao:2024,Elsherif:2022}. Although experimentally demonstrated in a~bent multimode fiber, the underlying statistical methodology is potentially applicable to other scattering environments, such as free-space optical channels affected by atmospheric turbulence~\cite{Klug:2023,Bachmann:2023}.

\section{\label{sec:methodology}Methodology}%

\begin{figure}[b!]%
	\includegraphics[width=\columnwidth]{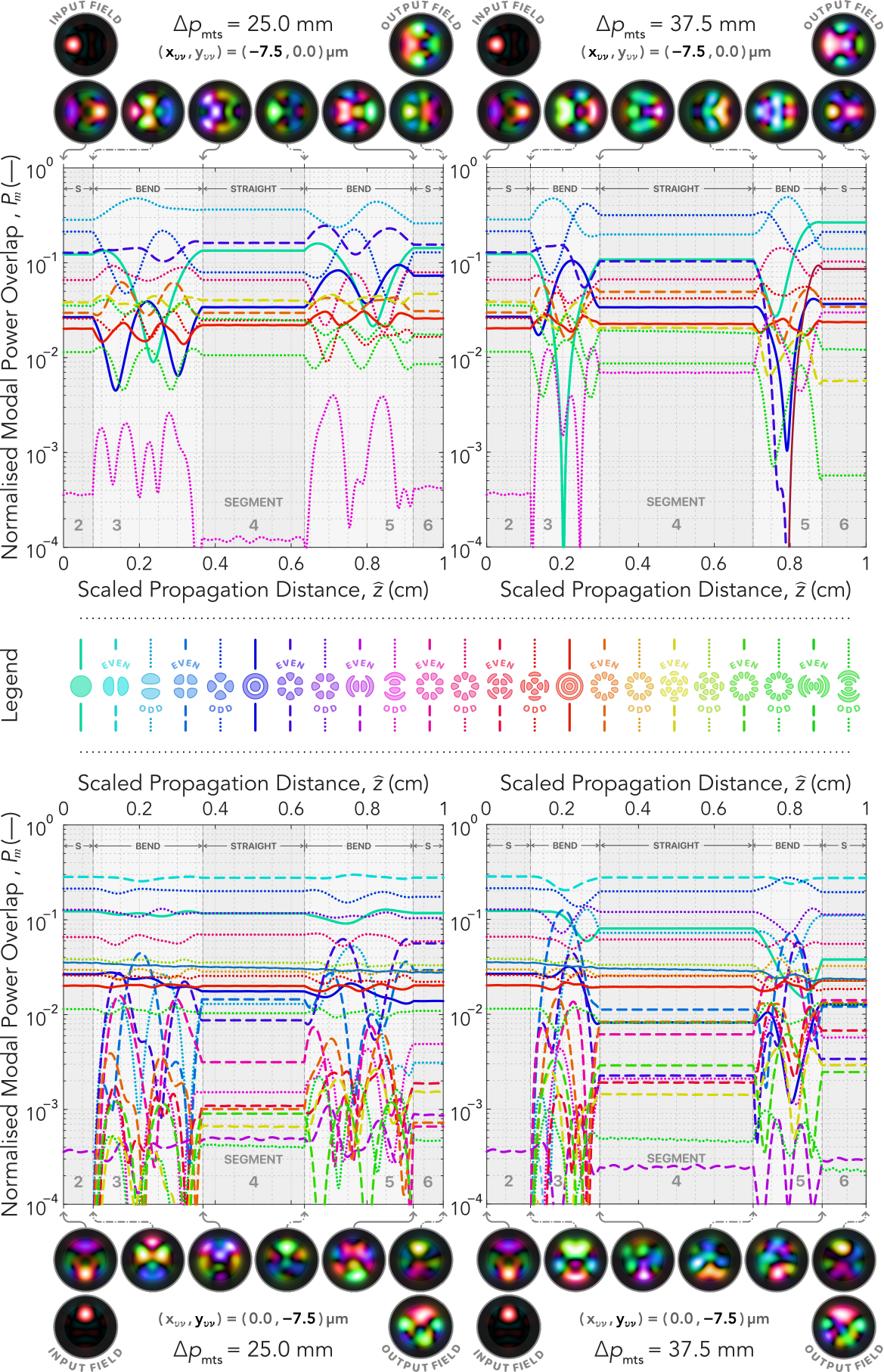}%
	\caption{\label{fig:simulations_PowerOverlap_allModes}Intermodal power exchange within the fiber’s modal basis is shown for an optical field propagating through a~U-shaped bend under linear deformations of $\Delta{p}_{\text{mts}} = \qtylist[list-open-bracket=\{,list-close-bracket=\},list-units=brackets,list-pair-separator={\text{,}\,},list-final-separator={\text{,}\,}]{25.0;37.5}{\milli\metre}$, corresponding to bend radii of  $r_\mathrm{B} = \qtylist[list-open-bracket=\{,list-close-bracket=\},list-units=brackets,list-pair-separator={\text{,}\,},list-final-separator={\text{,}\,}]{79.3;50.1}{\milli\metre}$. Symmetry preservation in the field is illustrated through spatial profiles derived from an input displaced  $\mathrm{LG}_{00}$ mode, defined as a~superposition of fiber modes at positions $\bigl\{(x_{\upsilon\nu},y_{\upsilon\nu})\bigr\} = \bigl\{(-\frac{3}{5}r_\text{core},0),(0,-\frac{3}{5}r_\text{core})\bigr\}$, where $r_\text{core} = \qty{12.5}{\micro\metre}$. Segmental lengths $L_s \mid s\in{\{2,\dots,6\}}$ scaled by the factor of $\num{0.01}\cdot{(\uppi R_{0})}^{-1}$, reducing the effective propagation distance to $\widehat{z}=\qty{1}{\centi\metre}$ to aid visualization of intermodal coupling.}%
	\end{figure}%
	
\begin{figure*}[t!]%
  \subcaptionbox{\label{fig:optical_setup}}{\includegraphics[width=0.71996419\textwidth]{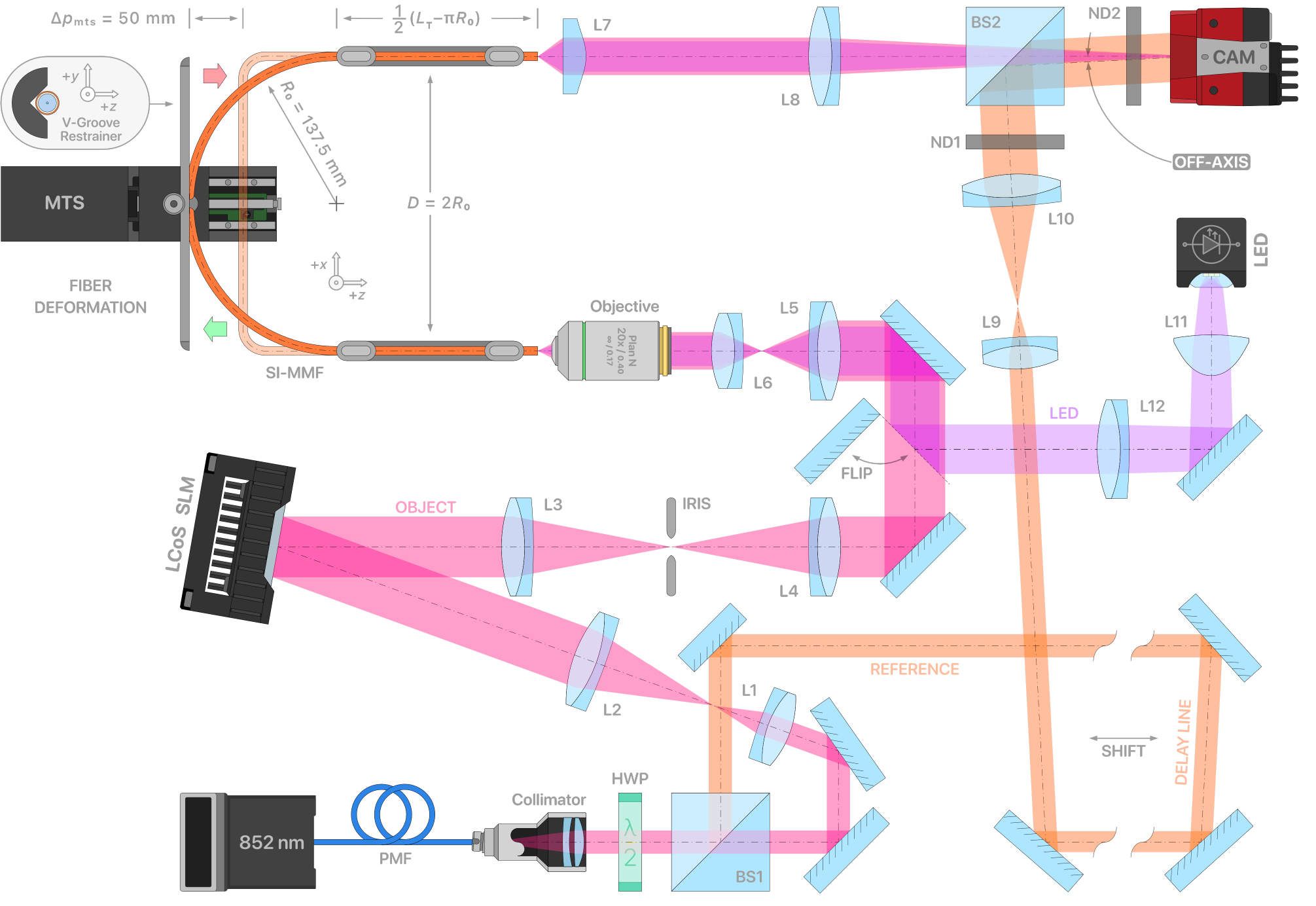}}\hfill%
  \subcaptionbox{\label{fig:geometrical_bend}}{\includegraphics[width=0.27681289\textwidth]{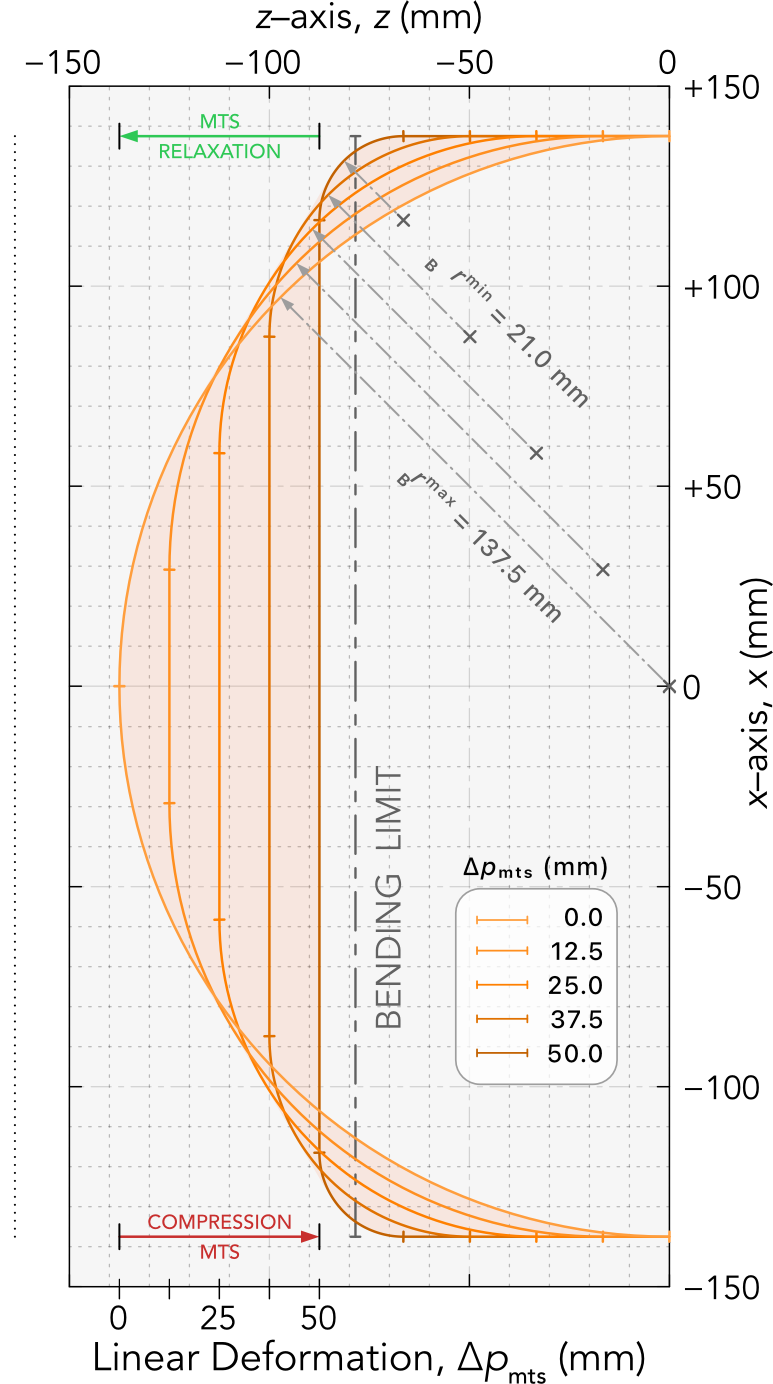}}%
	\caption{\label{fig:setup_bend}Measurement system for tracking mode field distortions in a~controllably bent fiber. (a)~Experimental optical setup enabling spatial scanning across the input facet of a~SI-MMF, repeated over a~series of controlled bending configurations imposed by the MTS. (b)~Idealized geometric model of the U-shaped fiber deformation, illustrating the predictable mechanical profile used to generate repeatable modal distortions. Fiber parameters: \ch{SiO2} core with $\dmtr_{\text{core}}=\qty{25(3)}{\micro\metre}$, \ch{F:SiO2} cladding with $\dmtr_{\text{cladd}}=\qty{125(2)}{\micro\metre}$, $\mathrm{N\!A} = \num{0.100(0.015)}$, $\alpha\approx\qty{0.048}{\decibel\per\metre}$ at $\lambda=\qty{850}{\nano\metre}$. Linear MTS is set to discrete positions $\Delta{p}_{\text{mts}}=\qtylist[list-open-bracket=\{,list-close-bracket=\},list-units=brackets,list-separator={,},list-final-separator={,\,\dots\,,}]{0.0;0.5;50.0}{\milli\metre}$ with a~positioning accuracy of $\overline{\delta}_{\text{mts}}=\pm\qty{1.6}{\micro\metre}$. Notes: BS -- beamsplitter, CAM -- camera, HWP -- half-wave plate, SI-MMF -- step-index multimode fiber, L -- lens, LCoS\,SLM -- liquid crystal on silicon spatial light modulator, LED -- light-emitting diode source at $\qty{850}{\nano\metre}$, MTS -- motorized translation stage, ND -- neutral-density filter, PMF -- polarization maintaining fiber.}%
\end{figure*}

\noindent In our approach, the fiber was controllably deformed into a~U-shaped configuration using a~flat V-groove fiber support, as illustrated in~\cref{fig:experimental_concept,fig:setup_bend,fig:fiberGeometry}. The support imposes a~repeatable planar deformation by modifying the radius of curvature of the fiber arc, as described in~\cref{sec:experimental_implementation}. These deformations can be approximated in simulation using a~small number of segments with known lengths and curvatures, based on the geometric model presented in~\cref{sec:bending_geometry,sec:transmission_throught_fiber}. See~\cref{tab:list_of_symbols} in \cref{sec:list_of_symbols} for a~complete list of quantities and symbols used throughout this manuscript. To assess the bending-induced modal effects, the output field was recorded under fixed input coupling while translating the bend longitudinally in the $xz$-plane. Speckle patterns at each position were decomposed via SVD into local orthonormal bases, which were then unified into a~single bend-resolved basis spanning the deformation-induced field subspace.%

\subsection{\label{sec:propagation_simulations}Propagation Simulations}

\noindent Simulations of the given fiber geometry were performed using an adapted finite-difference beam propagation method (FD-BPM), as described in~\cite{Veettikazhy:2021}. The results show that the spatial symmetry of the electric field at the distal end of the fiber $E(r, \phi, z = L_{\mathrm{T}})$, where $L_{\mathrm{T}}$ is the  total length of the fiber, is highly sensitive to the displacement of the diffraction-limited input spot on the proximal facet $E(r, \phi, z = 0)$, relative to the direction of curvature with respect to the fiber’s longitudinal axis. This is illustrated in~\cref{fig:simulations_PowerOverlap_allModes}, which presents normalized modal power overlap for two deformation levels $\Delta{p}_{\text{mts}} = \qtylist[list-open-bracket=\{,list-close-bracket=\},list-units=brackets,list-pair-separator={\text{,}\,},list-final-separator={\text{,}\,}]{25.0;37.5}{\milli\metre}$, along with input and output fields for segments $s\in\bigl\{2,3,4,5,6\bigr\}$. The associated bend radii are strictly given by~\cref{eq:bend_radii} as $r_\mathrm{B} = \qtylist[list-open-bracket=\{,list-close-bracket=\},list-units=brackets,list-pair-separator={\text{,}\,},list-final-separator={\text{,}\,}]{79.3;50.1}{\milli\metre}$, while the segment lengths $L_{s}$ are scaled by a~factor $\num{0.01} \cdot (\pi R_0)^{-1}$ to reduce the effective propagation distance to $\widehat{z} = \qty{1}{\centi\metre}$ for visual clarity of intermodal coupling. The propagating field preserves high symmetry when the bending plane aligns with the displacement of the input field relative to the center of the proximal fiber entrance, as demonstrated by the output fields at the top of~\cref{fig:simulations_PowerOverlap_allModes} for a~launched $\mathrm{LG}_{00}$ mode displaced by $(x_{\upsilon\nu},y_{\upsilon\nu}) = \qtylist[list-open-bracket=(,list-close-bracket=),list-units=brackets,list-pair-separator={\text{,}\,},list-final-separator={\text{,}\,}]{-7.5;0.0}{\micro\metre}$, where $(x_{\upsilon\nu}, y_{\upsilon\nu})\in\mathbb{R}^{2}$ denotes the transverse position of the launched beam center relative to the fiber core center. In contrast, when the bending plane is perpendicular to the input field displacement at the proximal facet, $(x_{\upsilon\nu},y_{\upsilon\nu}) = \qtylist[list-open-bracket=(,list-close-bracket=),list-units=brackets,list-pair-separator={\text{,}\,},list-final-separator={\text{,}\,}]{0.0;-7.5}{\micro\metre}$,  the resulting modal deformation disrupts this symmetry and induces pronounced asymmetry in the spatial power distribution, as demonstrated by the bottom axially unbalanced output fields. A~visual comparison of the top and bottom plots in~\cref{fig:simulations_PowerOverlap_allModes} shows that the nonlinear modal power coupling becomes significantly more intricate in the second scenario. 


\begin{figure}[t!]%
	\includegraphics[width=\columnwidth]{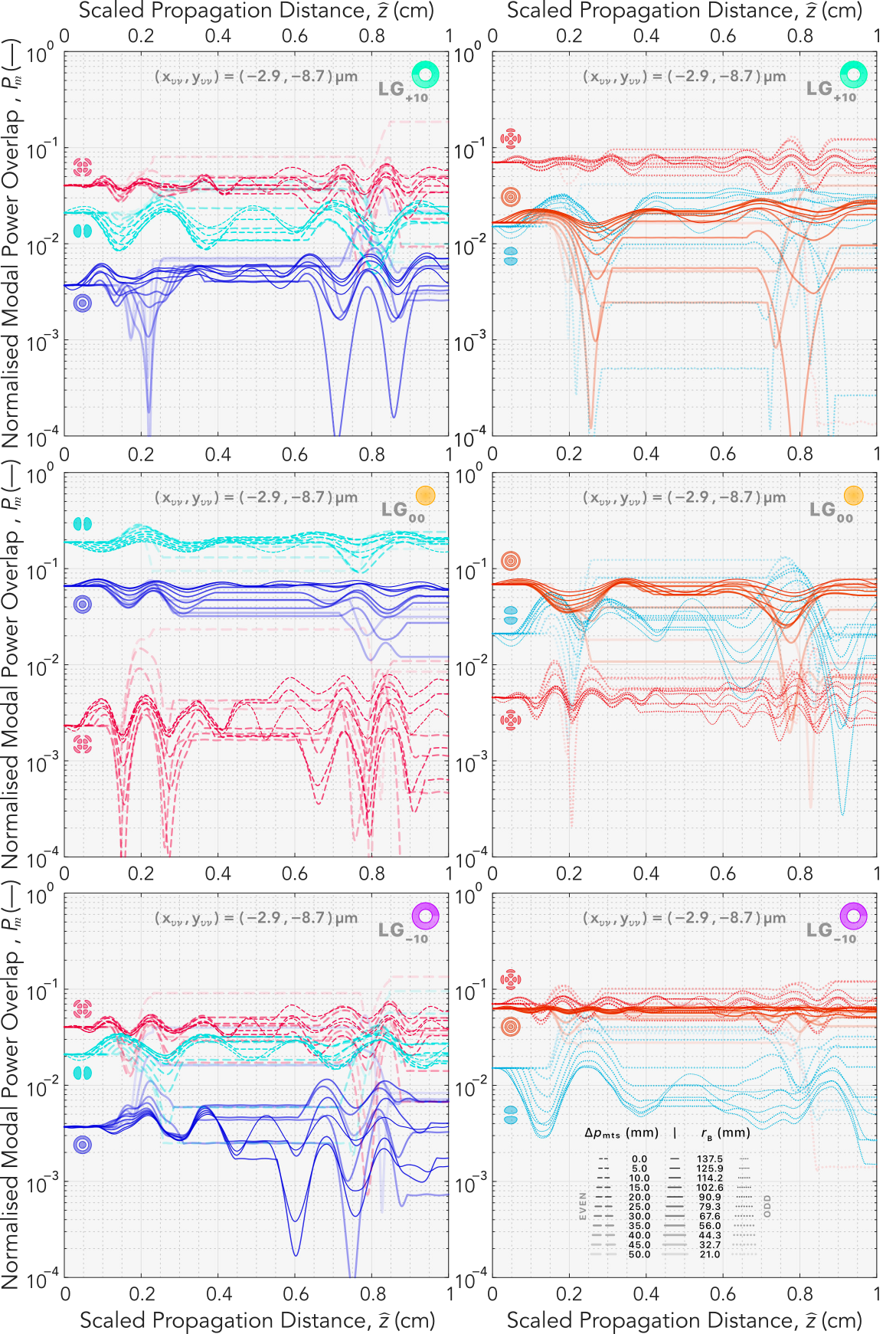}%
	\caption{\label{fig:simulations_PowerOverlap_fewPositions}Normalized modal power overlap for selected fiber modes $\big\{\mathrm{LP}_{\ell n}\mid(\ell,n)\in\{(1,1),(0,2),(2,2),(0,3)\}\big\}$, shown as a~function of linear deformation values $\Delta{p}_{\text{mts}} = \qtylist[list-open-bracket=\{,list-close-bracket=\},list-units=brackets,list-separator={,},list-final-separator={,\,\dots\,,}]{0.0;5.0;10.0;50.0}{\milli\metre}$, corresponding to bend radii $r_\mathrm{B} = \qtylist[list-open-bracket=\{,list-close-bracket=\},list-units=brackets,list-separator={,},list-final-separator={,\,\dots\,,}]{137.5;125.9;114.2;21.0}{\milli\metre}$. The configuration involves a~U-shaped bend and OAM-carrying inputs $\mathrm{LG}_{\ell 0}\mid\ell\in\{\pm 1,0\}$, launched at a~fixed entrance position $(x_{\upsilon\nu}, y_{\upsilon\nu}) = \left( -\tfrac{3}{13}r_\text{core},\ -\tfrac{9}{13}r_\text{core} \right)$, where $r_\text{core} = \qty{12.5}{\micro\metre}$. Segmental lengths $L_{s}\mid{s}\in \{2,\dots,6\}$ are scaled by a~factor $\num{0.01}\cdot(\uppi R_{0})^{-1}$, reducing the effective propagation distance to $\widehat{z} = \qty{1}{\centi\metre}$ to enhance visual clarity in the illustrated power dynamics.}%
	\end{figure}%
	

Similarly, additional results in~\cref{fig:simulations_PowerOverlap_fewPositions} depict intermodal power exchange for distinct OAM-coupling scenarios involving modes $\big\{\mathrm{LG}_{\ell 0}\mid \ell \in \{-1, 0, +1\}\big\}$, launched at a~fixed fiber entrance position $(x_{\upsilon\nu},y_{\upsilon\nu}) = (-2.9,-8.7)\,\unit{\micro\metre}$, as shown in~\cref{fig:experimental_concept}. The left and right columns of~\cref{fig:simulations_PowerOverlap_fewPositions} depict the normalized power overlap for the fundamental modes $\mathrm{LP}_{02}$ and $\mathrm{LP}{03}$, respectively, shown alongside the even modes $\mathrm{LP}_{11}$, $\mathrm{LP}_{22}$ and their odd counterparts, subjected to discrete linear deformations in the range $\Delta{p}_\text{mts}\in\qtyrange[range-units=brackets,range-phrase=\text{--}]{0}{50}{\milli\metre}$ with a~step size of $\delta{p}_{\text{mts}}=\qty{5}{\milli\metre}$, corresponding to bend radii in the range ${r}_{\mathrm{B}}=\qtyrange[range-units=brackets,range-phrase=\text{--}]{137.5}{21.0}{\milli\metre}$ with a~step size of $\delta{r}_{\mathrm{B}}\approx\qty{11.6}{\milli\metre}$. The emerging beam, after propagating through a~fiber curvature oriented in an arbitrary plane, reveals that phase delays induced by modal conformations tend to localize intermodal energy exchange within the plane of the bend. It should be noted, however, that these observations rely on an idealized model where the curvature is restricted to a~single longitudinal plane, the transverse refractive index deformation is uniform, micro-bends are neglected, and the twist rate asymptotically approaches zero over an infinitely long propagation path.%

\subsection{\label{sec:experimental_measurements}Experimental Measurements}

\noindent The complex wavefront of the speckle field emerging at the distal facet of the fiber was measured using off-axis digital holography, following excitation by a~sequence of spatially displaced, diffraction-limited focal spots at the proximal facet. These focal positions were arranged on a~square Cartesian grid in the Fourier plane, defined as $\left\{ (x_{\upsilon\nu}, y_{\upsilon\nu}) \right\} = \lambda f_{\mathrm{L}} d^{-1} \cdot \left\{ (\upsilon, \nu) \;\middle|\; (\upsilon, \nu) \in \mathbb{Z}^2 \right\}$, where $(\upsilon, \nu)$ are integer index pairs corresponding to discrete linear phase ramps applied to the SLM to deflect the first diffraction order, $f_{\mathrm{L}}$ denotes the effective focal length of the lens system preceding the fiber, and $d = {n}_{\text{px}}\cdot{p}_{\text{SLM}}$ is the grating period, determined by the SLM pixel pitch and the number of pixels per phase ramp. In addition to performing a~spatial sweep using a~Gaussian beam $\mathrm{LG}_{00}$ to excite a~range of fiber modes and evaluate sensitivity to bending under varying launch conditions, further experiments were conducted with beams carrying orbital angular momentum (OAM), specifically the higher-order modes $\mathrm{LG}_{\pm 10}$. The notation adopted in the following analysis is independent of the specific input mode, as the resulting output fields corresponding to each injected beam were processed separately for each displacement, assuming a~fixed $\mathrm{LG}$ mode during each acquisition. In this manner, for each linear deformation position $\Delta{p}_{\text{mts}}=\qtylist[list-open-bracket=\{,list-close-bracket=\},list-units=brackets,list-separator={,},list-final-separator={,\,\dots\,,}]{0.0;0.5;50.0}{\milli\metre}$ imposed by the motorized translation stage (MTS), light was injected into the proximal facet of the fiber for input modes indexed by $\upsilon,\nu\in\llbracket{-13},{+13}\rrbracket$, as illustrated in~\cref{fig:volumetricSpeckle_Fiber}. Axial coupling at the core center $(\upsilon,\nu)=(0, 0)$ predominantly excites fundamental modes, whereas increasing the eccentric displacement of the injection spot facilitates power coupling into higher-order modes, albeit with lower efficiency near the core--cladding periphery. Since the displacement of the input beam selectively excites a~subset of modes with distinct power distributions, the resulting intermodal coupling, and therefore the fiber’s bending response, is inherently dependent on the launch position. Accounting for this spatial dependence is critical for experimentally resolving bending-induced modal evolution, with dense input sampling enabling indirect inference of intermodal power redistribution from speckle evolution, as described in~\cref{sec:bending_matrices}, for correlative analysis under fixed coupling with the results depicted in~\cref{fig:innerProductsBend_All}.

The measured complex electric fields forming speckle patterns at the distal facet, resulting from various input displacements and MTS-induced deformations, are denoted by $S_{\mu}^{(p)}\!\left(r,\theta\right) = E_{\mu}^{(p)}\!\left(r,\theta,z=L_{\mathrm{T}}\right)$, where the global index $\mu = \mu(\upsilon,\nu)$ maps each input coordinate pair $(\upsilon,\nu)\in\mathbb{Z}^2$ to a~unique integer $\mu\in\mathbb{N}$, and $p\in\mathbb{N}$ indexes the discrete deformation positions imposed along the fiber by the motorized translation stage $\Delta p_{\text{mts}} = p\cdot\delta{p}_\text{mts}$, where $\delta{p}_{\text{mts}} \in \mathbb{R}^{+}$ is a~uniform step size. A~total of $N_\mathrm{S} = 729$ complex speckle fields were acquired across $N_\mathrm{B} = 101$ discrete bending configurations, resulting from spatially displaced excitations scanned over the proximal facet. Although the number of input positions significantly exceeds the number of supported guided modes, $N_{\mathrm{M}} \approx \num{23}$, this deliberate oversampling ensures systematic variation in the launched modal superpositions. Each displacement excites an overlapping subset of modes with slightly different angular and radial weightings. The resulting spatial diversity enhances sensitivity to deformation-induced modal dynamics and enables robust experimental characterization of intermodal coupling by capturing bending-dependent variations in the output fields across a~densely sampled excitation grid.


\subsection{\label{sec:singular_value_decomposition}Singular Value Decomposition}
\noindent As the propagation dynamics are examined across a~dense set of lateral displacements at the fiber input, the resulting dataset enables continuous tracking of deformation-induced power exchange. However, the total information captured in this speckle space, spanned by the measured output fields, is highly redundant and overestimated, since the number of supported guided modes $N_\mathrm{M} \ll N_\mathrm{S}$, where $N_\mathrm{S}$ is the number of recorded speckle patterns and $N_\mathrm{M}$ is the number of propagating modes supported by the fiber. Although a~sufficiently large set of speckle fields can represent arbitrary outputs as linear combinations, the resulting basis is overcomplete and non-orthogonal, i.e., $\braket{S_{\mu}}{S_{\eta}}\neq{0}$ for $\mu\neq\eta$, where $\eta = \mu^{\prime}$ is explicitly used to distinguish the second index for clarity, and $\braket{\cdot}{\cdot}$ denotes the inner product written in bra--ket notation. Despite the dimensional mismatch, scanning with $N_\mathrm{S} \gg N_\mathrm{M}$ is essential to probe the full modal subspace and capture subtle deformation-induced dynamics. To extract a~compact and physically meaningful representation, we apply SVD to reduce the dimensionality while retaining dominant power-exchanging features within the fiber’s modal space~\cite{Kupianskyi:2024}. For each discrete deformation state indexed by $p\in\mathbb{N}$, the corresponding set of measured output fields under constant fiber deformation $\bigl\{S_{1}^{(p)},S_{2}^{(p)},\dots,S_{N_{\mathrm{S}}}^{(p)}\big\}$ is used to construct a~bend-dependent correlation matrix $\mat{M}^{(p)} \in \mathbb{C}^{N_\mathrm{S} \times N_\mathrm{S}}$  defined through a~complex-valued overlap integral as
\begin{equation}\label{eq:inner_product}
	M_{\mu\eta}^{(p)} = \int\limits_{-\uppi}^{+\uppi}\!\int\limits_{0}^{\infty}\left({S_{\mu}^{(p)}\!\left(r,\theta\right)}\right)^{\ast}\,S_{\eta}^{(p)}\!\left(r,\theta\right)\;r\odif{r}\odif{\theta},
\end{equation}
where the elements $M_{\mu\eta}^{(p)} = \braket{S_{\mu}^{(p)}}{S_{\eta}^{(p)}}$ define the Hermitian matrix $\mathbf{M}^{(p)}$, satisfying $M_{\mu\eta}^{(p)} = \big(M_{\eta\mu}^{(p)}\big)^{\ast}$. To elucidate the intrinsic modal dynamics and interactions within the fiber, these matrices are subjected to SVD,  enabling the extraction of linearly independent, orthogonal components that capture the dominant power distribution while effectively spanning the fiber’s guided modal space~\cite{Miller:2019}. As a~result, each correlation matrix $\mat{M}^{(p)}$ associated with a~discrete fiber deformation state can be expressed through singular value decomposition in its general form as
\begin{equation}\label{eq:svd}
\mat{M}^{(p)} = \mat{U}^{(p)}\mat{\Sigma}^{(p)}\big(\mat{V}^{(p)}\big)^{\mathsf{H}} = \sum_{\mathcal{k}=1}^{N_{\mathrm{S}}} \sigma_{\mathcal{k}}^{(p)} \vek{u}_{\mathcal{k}}^{(p)} \big(\vek{v}_{\mathcal{k}}^{(p)}\big)^{\mathsf{H}},
\end{equation}
where $\mat{U}^{(p)}, \mat{V}^{(p)} \in \mathbb{C}^{N_\mathrm{S} \times N_\mathrm{S}}$ are unitary matrices whose columns $\vek{u}_{\mathcal{k}}^{(p)}, \vek{v}_{\mathcal{k}}^{(p)} \in \mathbb{C}^{N_\mathrm{S}}$ represent the left and right singular vectors, respectively, and $\mat{\Sigma}^{(p)} \in \mathbb{R}^{N_\mathrm{S} \times N_\mathrm{S}}$ is a~diagonal matrix whose non-negative real entries $\sigma_{\mathcal{k}}^{(p)} \in \mathbb{R}$ denote the singular values, indexed by $\mathcal{k} \in \mathbb{N}$ and arranged in descending order of magnitude. For notational clarity, the superscript $(p)$ is henceforth omitted, with all quantities understood to depend implicitly on the discrete deformation state $\Delta p_{\text{mts}}\in\qtyrange[range-units=brackets,range-phrase=\text{--}]{0}{50}{\milli\metre}$. 

Since the correlation matrix Hermitian, satisfying $\mat{M} = \mat{M}^{\mathsf{H}} = (\mat{U} \mat{\Sigma} \mat{V}^{\mathsf{H}})^{\mathsf{H}} = \mat{V} \mat{\Sigma} \mat{U}^{\mathsf{H}}$, it follows that $\mat{V} = \mat{U}$.  As a~result, the SVD of $\mat{M}$ reduces to its eigendecomposition, $\mat{M} = \mat{V} \mat{\Lambda} \mat{V}^{\mathsf{H}}$, where $\sigma_{\mathcal{k}}$ are equivalent to the square roots of the eigenvalues $(\lambda_{\mathcal{k}})$, since $\mat{\Lambda} = \mat{\Sigma}^2$. Thus the each SVD can be interpreted as an eigendecomposition of overdetermined Hermitian matrix expressed as $\mat{M} = \mat{V} \mat{\Lambda} \mat{V}^{\mathsf{H}}$, where $\sigma_{\mathcal{k}}$ are equivalent to the square roots of the eigenvalues $(\lambda_{\mathcal{k}})$, since $\mat{\Lambda} = \mat{\Sigma}^2$. Consequently, the diagonal of $\mat{\Sigma}^2 = \mat{V}^{\mathsf{H}} \mat{M} \mat{V}$ contains the eigenvalues of $\mat{M}$, and the columns of $\mat{V}$ are the corresponding eigenvectors. As the singular values in each overcomplete non-unitary matrix (of rank $\mathcal{r} = N_{\mathrm{S}}$) are arranged in descending order as $\left|\sigma_{\mathcal{k}}\right|^{2}\geq\left|\sigma_{\mathcal{k}+1}\right|^{2}\geq0\;\forall\;\mathcal{k}\!\in\left\{1,2,\dots,N_{\mathrm{S}}\right\}$, it should be possible to identify a~cutoff threshold ($\epsilon>0$) in the singular value spectrum that determines the effective rank ($\mathcal{r}_\mathrm{X}$) of the matrix $\mat{M}$ equivalent corresponding to the fiber modal space as ’s modal space as $\text{rk}(\mat{M})\equiv{N_{\mathrm{M}}}$, such that $\left|\sigma_{\mathcal{k}}\right|^{2}\leq\epsilon\;\forall\;\mathcal{k}>\mathcal{r}_\mathrm{X}$ and $\left|\sigma_{\mathcal{k}}\right|^{2}\gg\epsilon\;\forall\;\mathcal{k}\leq\mathcal{r}_\mathrm{X}$. In an ideal scenario devoid of noise or measurement artifacts, the true rank of the system precisely matches the number of guided modes, such that $\mathcal{r}_{\mathrm{X}} = N_{\mathrm{M}}$. The presence of noise or experimental imperfections may artificially increase the apparent rank, leading to $\mathcal{r}_{\widetilde{\mathrm{X}}} > N_{\mathrm{M}}$ and obscuring the true modal bandwidth of the fiber.%


\section{Results}

\begin{figure*}[t!]%
  \subcaptionbox{\label{fig:modesSVD}$\Psi_{\eta}^{(p)}\leftarrow\mathrm{LG}_{00}$}{\includegraphics[width=0.711\textwidth]{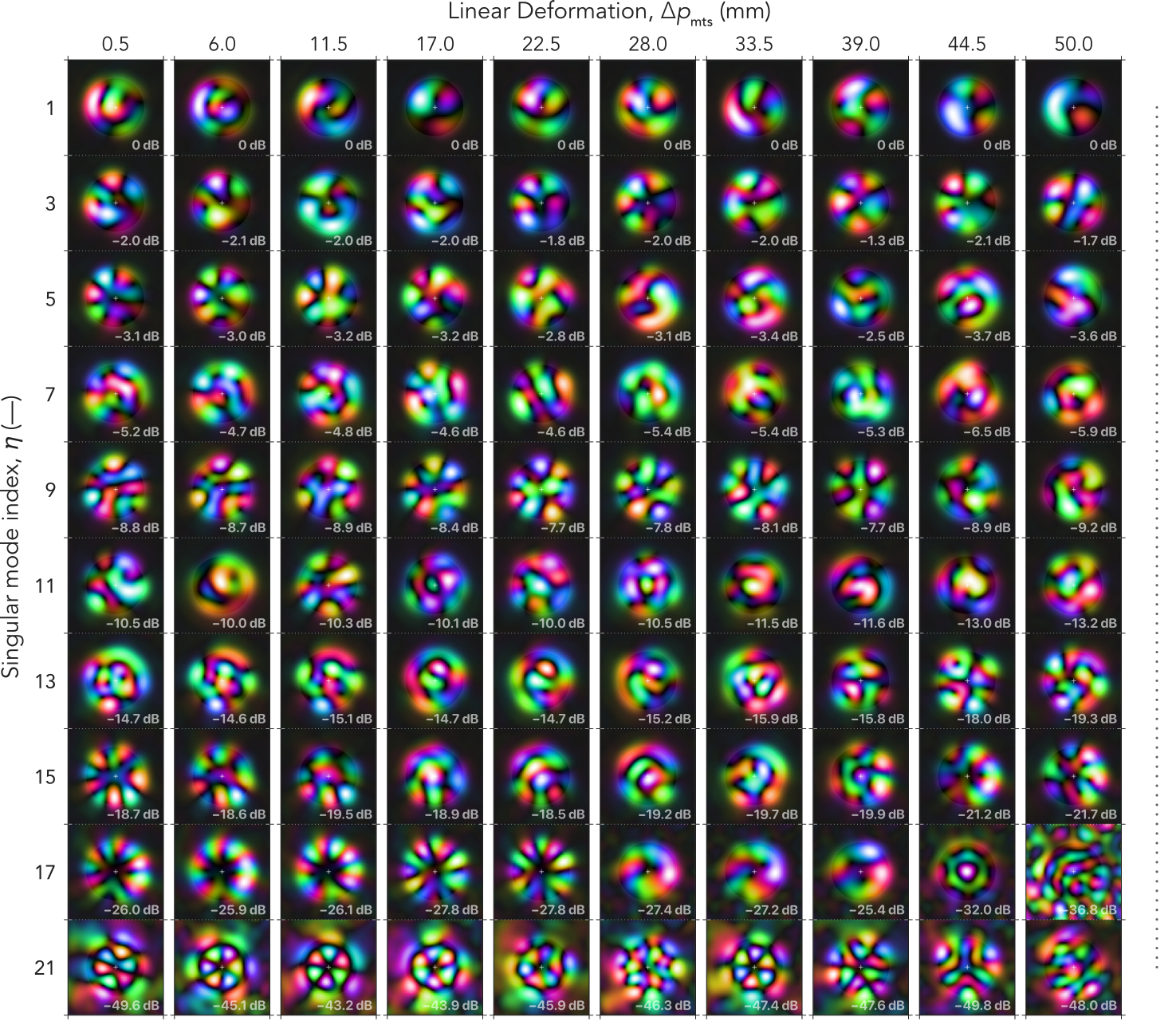}}\hfill%
  \subcaptionbox{\label{fig:couplingStrengthsSVD}$\sigma_{\eta}^{(p)}\leftarrow\mathrm{LG}_{\{00,\pm10\}}$}{\includegraphics[width=0.289\textwidth]{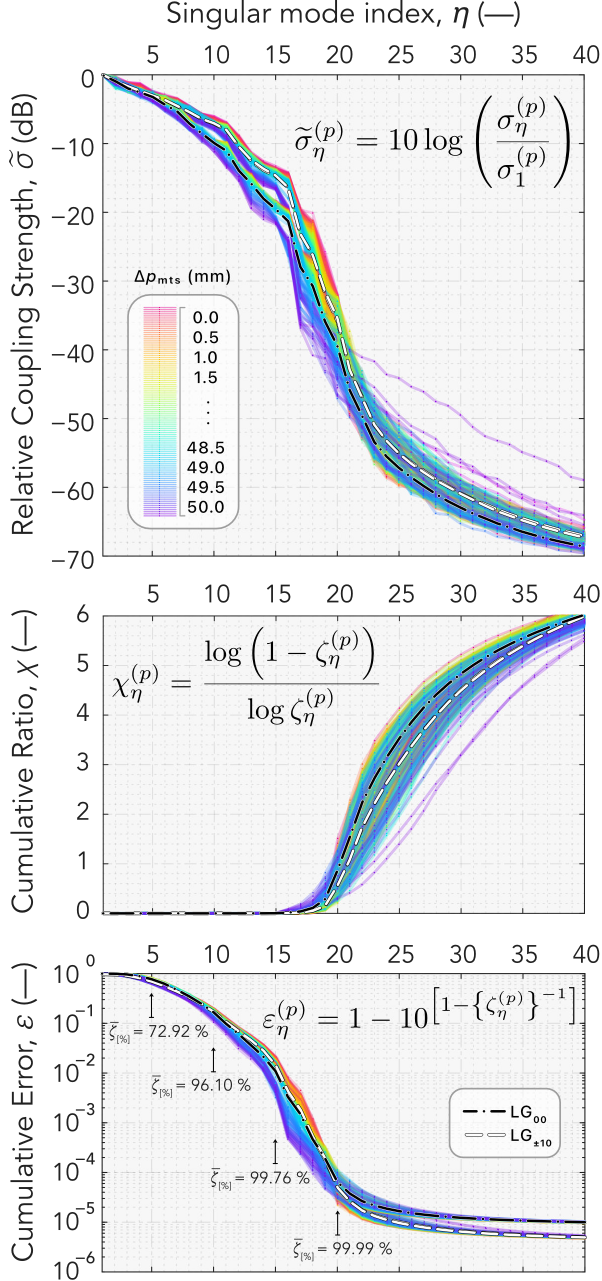}}%
	\caption{\label{fig:modesSVD_couplingStrengthsSVD}An overview of the experimentally constructed deformation-resolved sets of singular modes, ordered by decreasing singular values. (a)~A plot illustrates the variation in the spatial profiles of singular modes, where $\braket{\Psi_{\eta}^{(p)}}{\Psi_{\eta^{\prime}}^{(p^{\prime})}} = \delta_{pp^{\prime}}\delta_{\eta\eta^{\prime}}$, under approximately equal coupling strengths as the fiber undergoes linear deformation via MTS translation. The modes are derived from speckle patterns generated through $\mathrm{LG_{00}}$ excitation. (b)~Relative coupling strengths for all linear deformations $\Delta{p}_\text{mts}$, visualized through plots of singular values derived from the cumulative energy content $\zeta_{\eta}^{(p)}\!\in[0,1]$, expressed as $\zeta^{(p)}_{\eta} = \sum_{i=1}^{\eta}\sigma_{i}^{(p)}\bigl(\sum_{i=1}^{N_\mathrm{S}}\sigma^{(p)}_{i}\bigr)^{-1}$. The average of the given quantities is indicated by the black dash-dotted line for $\mathrm{LG_{00}}$ and the white dashed line for $\mathrm{LG_{\pm10}}$, respectively.}%
\end{figure*}%


\begin{figure*}[t!]%
  \subcaptionbox{\label{fig:innerProductSVD_4_LG-10}$\eta=4\leftarrow\mathrm{LG}_{-10}$}{\includegraphics[width=0.259423036\textwidth]{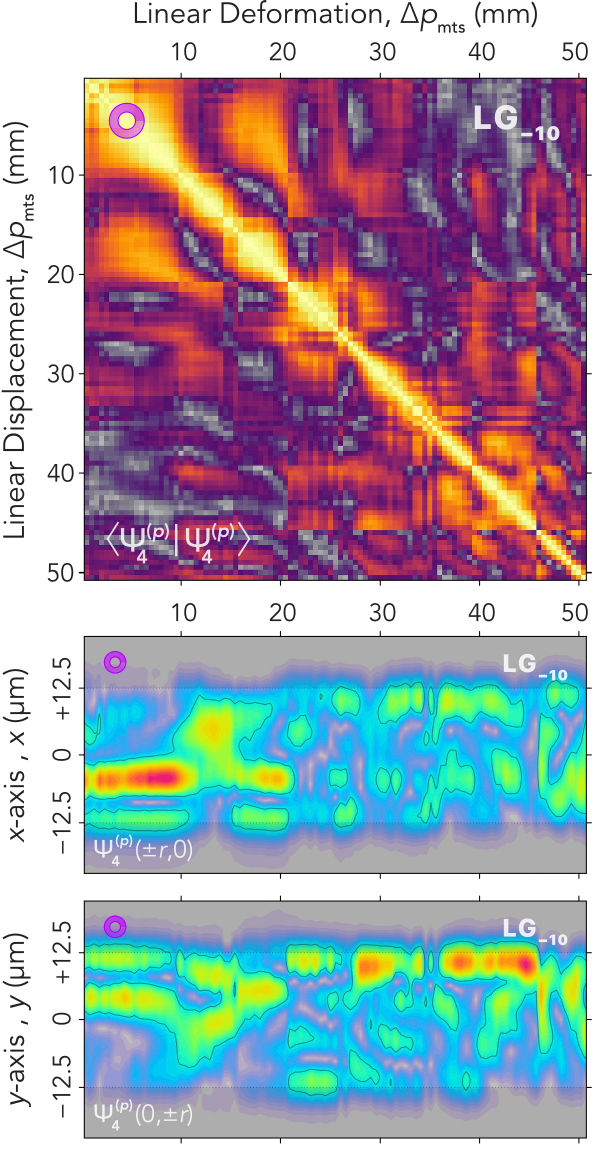}}
  \subcaptionbox{\label{fig:innerProductSVD_4_LG00}$\eta=4\leftarrow\mathrm{LG}_{00}$}{\includegraphics[width=0.236260265\textwidth]{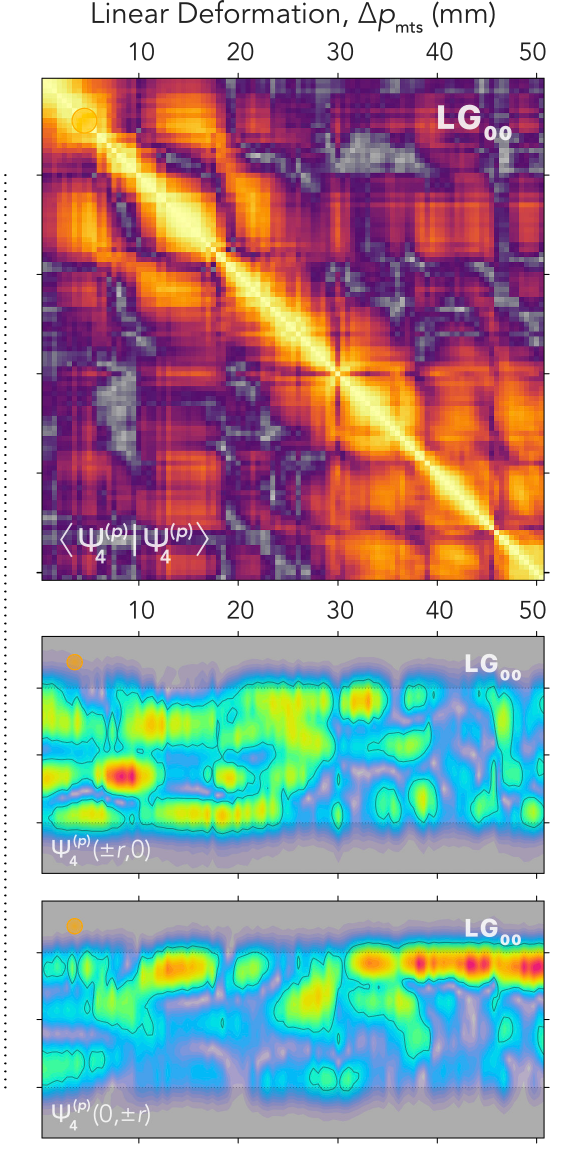}}
    \subcaptionbox{\label{fig:innerProductSVD_7_LG00}$\eta=7\leftarrow\mathrm{LG}_{00}$}{\includegraphics[width=0.236260265\textwidth]{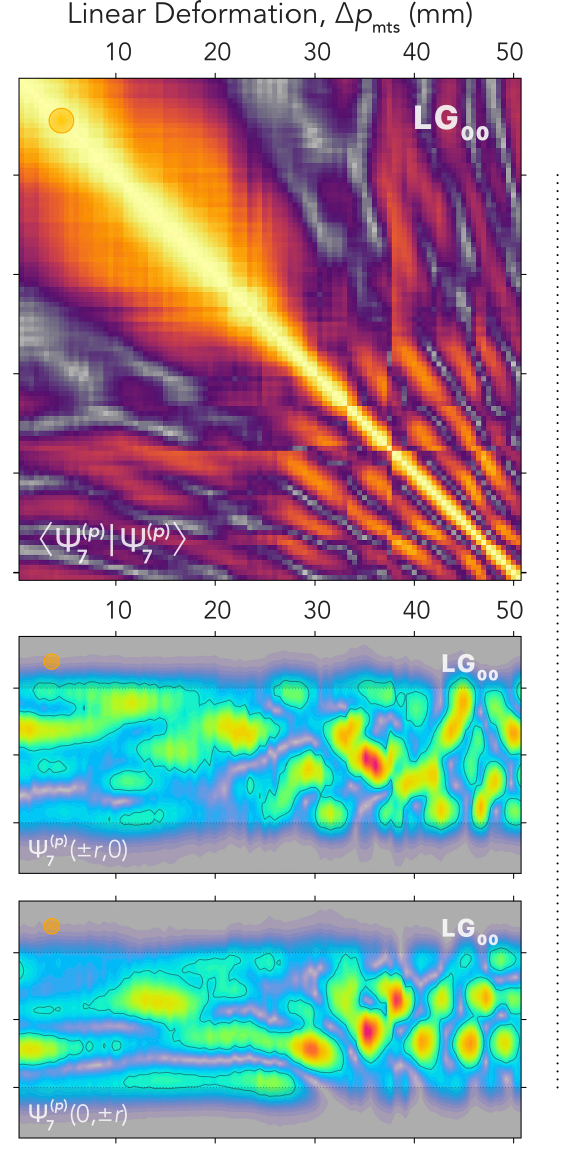}}
      \subcaptionbox{$\eta=7\leftarrow\mathrm{LG}_{+10}$\label{fig:innerProductSVD_7_LG+10}}{\includegraphics[width=0.268056433\textwidth]{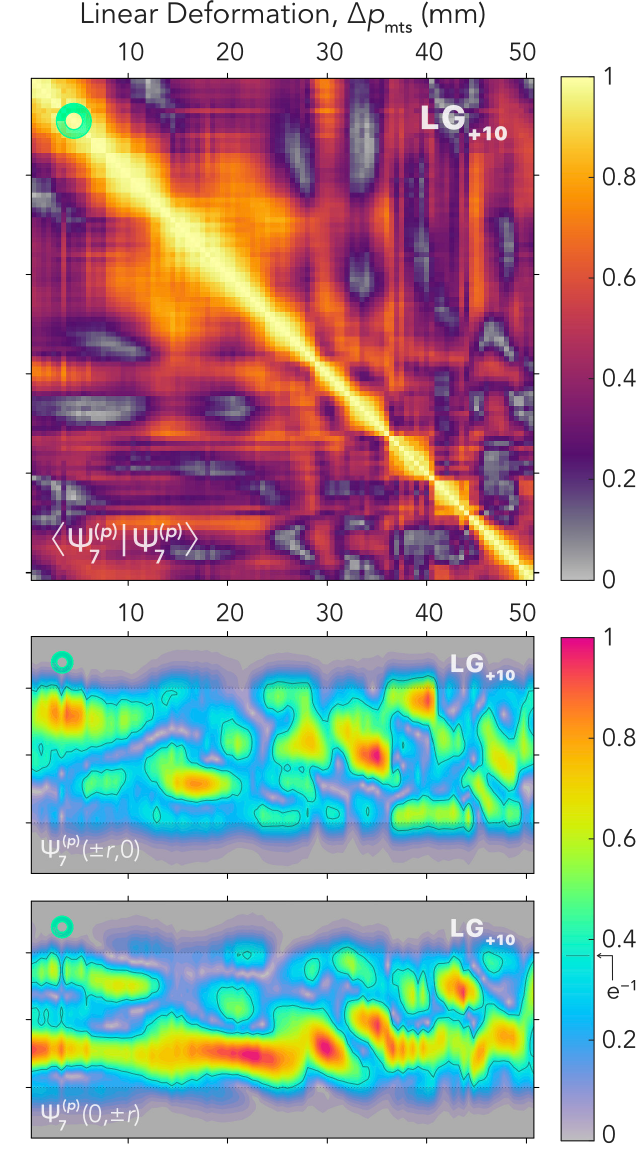}}
	\caption{\label{fig:innerProductSVD}Correlation between experimentally retrieved singular modes with comparable coupling strengths across all linear deformations,  where modes originating from different deformation states lack orthogonality, $\braket{\Psi_{\eta}^{(p)}}{\Psi_{\eta}^{(p^{\prime})}}\neq{0}$ for ${p\neq{p}^{\prime}}$. Spatial variation of singular modes under $\mathrm{LG}_{00}$ and $\mathrm{LG}_{\pm10}$, excitation leads to distinct correlation matrices; compare (a,\,b) $\eta={4}$ and (c,\,d) $\eta={7}$. Corresponding field cross-sections $\big|{\Psi^{(p)}_{\eta}\!(x_0,y_0)}\big|$ are shown for $(x_0, y_0) = \bigl\{(\pm{r}, 0), (0,\pm{r})\bigr\}$, remaining strictly orthogonal  in (b,\,c) across all deformations $\Delta{p}_{\text{mts}}=\qtylist[list-open-bracket=\{,list-close-bracket=\},list-units=brackets,list-separator={,},list-final-separator={,\,\dots\,,}]{0.0;0.5;50.0}{\milli\metre}$.}%
\end{figure*}%


\begin{figure*}[t!]%
  \subcaptionbox{\label{fig:robustSVD_LG-10}$\widehat{\Psi}_{\rho}\leftarrow\mathrm{LG}_{-10}$}{\includegraphics[width=0.18416163\textwidth]{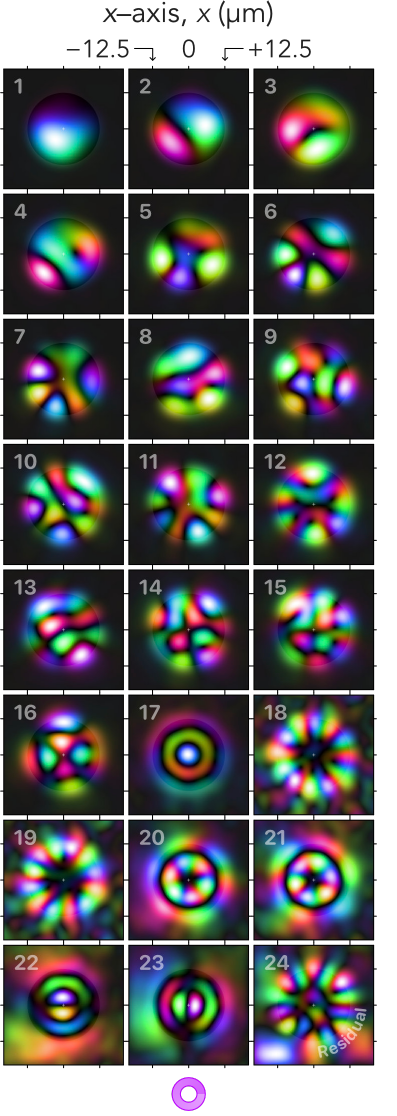}}\hfill%
  \subcaptionbox{\label{fig:robustSVD_LG00}$\widehat{\Psi}_{\rho}\leftarrow\mathrm{LG}_{00}$}{\includegraphics[width=0.63167673\textwidth]{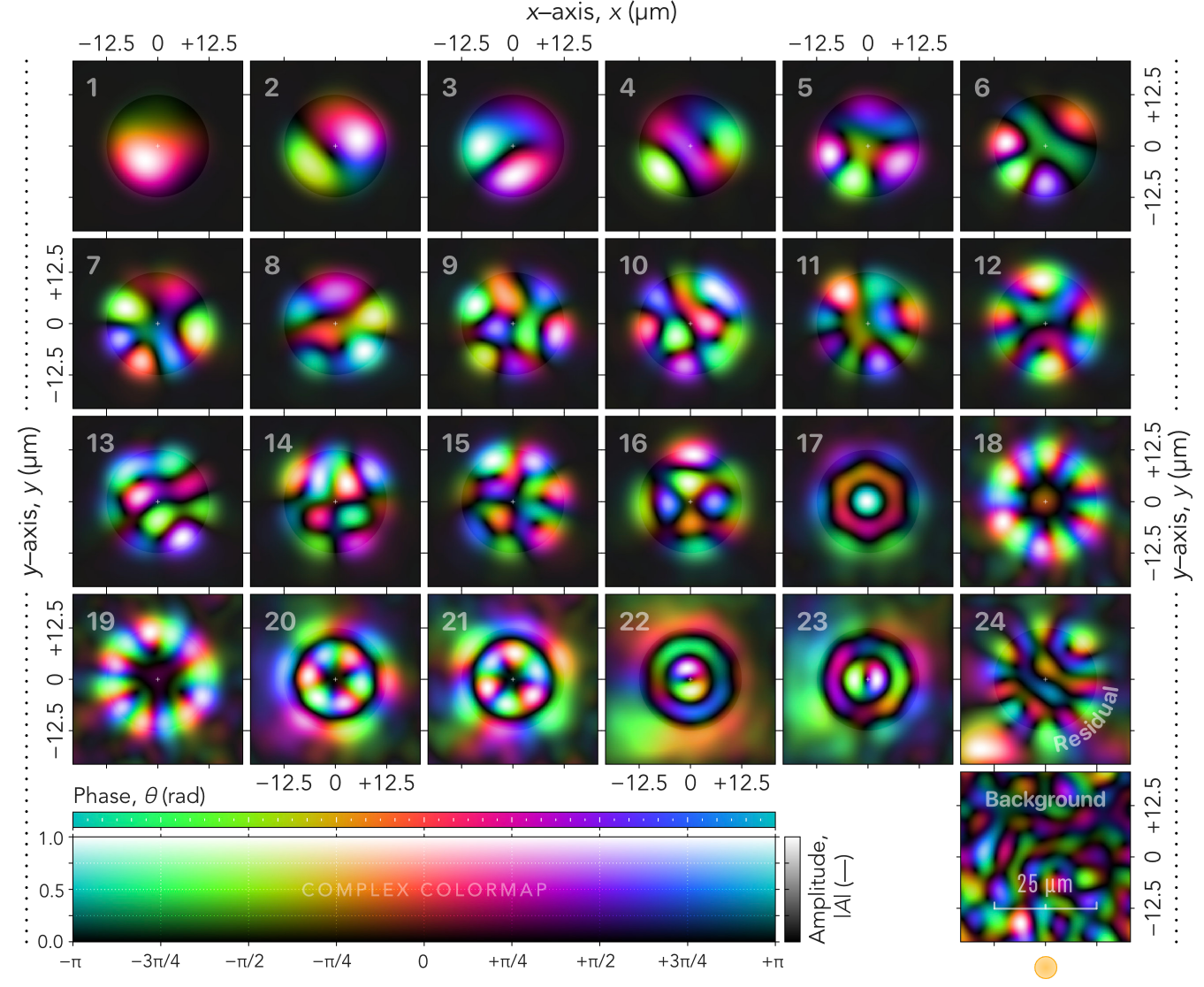}}\hfill%
    \subcaptionbox{\label{fig:robustSVD_LG+10}$\widehat{\Psi}_{\rho}\leftarrow\mathrm{LG}_{+10}$}{\includegraphics[width=0.18416163\textwidth]{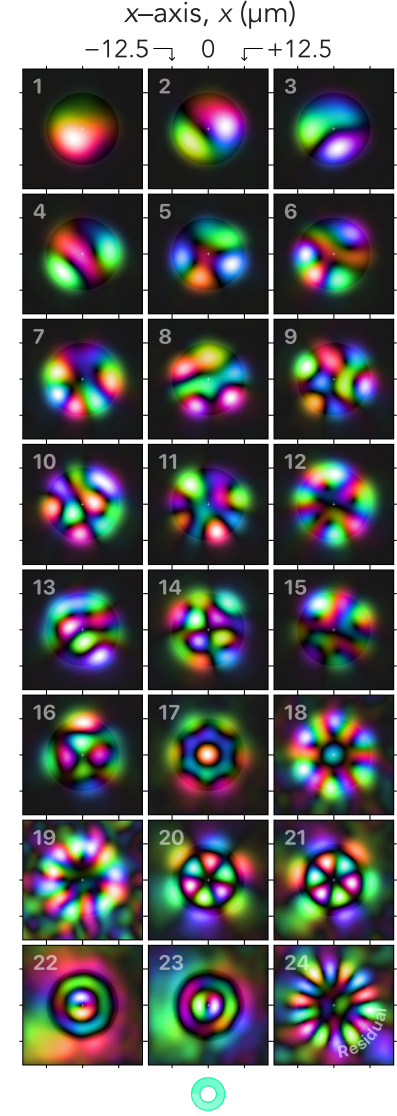}}
	\caption{\label{fig:robustSVD}A~deformation-resolved global basis that provides a~compact, physically meaningful representation of the modal subspace induced by mechanical deformations. Unlike the initial per-deformation SVD, which yields locally orthonormal but unaligned bases, the second SVD reorients and consolidates them into a~unified coordinate system aligned with the empirical structure of modal evolution under bending. (a,\,b,\,c) Excitation with $\mathrm{LG}_{\ell0}\mid\ell\in\{{-1,0,+1}\}$ yields consistent spatial profiles within the global basis regardless of coupling conditions, in contrast to the deformation-sensitive patterns observed in the initial SVD decomposition. Field profiles are visualized using a complex-valued colormap, with the amplitude normalized to the maximum value of each mode profile.}%
\end{figure*}%


\noindent To rigorously characterize the fiber’s optical response under specific deformations, we construct a set of compact, orthonormal bases that capture the underlying modal structure solely from experimentally measured speckle statistics, as described in~\cref{sec:experimental_measurements}, by leveraging the per-deformation modal decomposition framework introduced in~\cref{sec:singular_value_decomposition}. For each discrete mechanical configuration imposed by the MTS, a~deformation-resolved correlation matrix is constructed from the ensemble of speckle fields recorded at the fiber output, and its SVD yields a~complete, finite, and orthonormal basis of effective modes tailored to that particular deformation, defined as
\begin{equation}\label{eq:construction_svd}
	{\Psi}_{\eta}^{(p)}\!\left(r,\theta \right) = \big(\sigma_{\eta}^{(p)}\big)^{-\frac{1}{2}} \sum_{\mu=1}^{N_\mathrm{S}} \big(V_{\mu\eta}^{(p)}\big)^{\ast}\,S_{\mu}^{(p)}\!(r,\theta),
\end{equation}
resulting in a set of mutually orthogonal and linearly independent modes. These singular modes represent weighted linear combinations of the fiber’s intrinsic modes, forming a~compact, low-rank approximate basis that spans the fiber’s modal space and can represent any arbitrary optical field emerging from the same fiber. For a~specific bend configuration, the singular modes within each set are inherently orthogonal, rigorously defined as $\braket{\Psi_{\eta}^{(p)}}{\Psi_{\eta^{\prime}}^{(p^{\prime})}} = \delta_{pp^{\prime}}\delta_{\eta\eta^{\prime}}$, thereby ensuring modal orthogonality within each deformation-resolved set and eliminating crosstalk between modes of the same configuration. The orthogonality originates from the SVD itself, wherein the singular value matrix, expressed as $\mat{V}^\mathsf{H} \mat{M} \mat{U} = \mat{\Sigma}$ is a~strictly diagonal and real matrix that ensures the resulting modes are linearly independent and mutually decoupled~\cite{Ho:2014}. The singular values can be interpreted as deformation-specific transmission coefficients that quantify the relative contribution of each modal component to the total output field, implicitly reflecting power redistribution and attenuation effects across the system. If the inverse square roots of these singular values are omitted in the normalization of \cref{eq:construction_svd}, the resulting basis remains orthogonal but no longer orthonormal, thereby preserving the relative power distribution among the fiber modes rather than equalizing their norms. Consequently, each per-deformation set of singular modes constitutes an optimal basis that faithfully represents the modal structure and enables accurate reconstruction of arbitrary output fields generated under the same fiber configuration, despite the basis itself not being invariant under fiber propagation.%


Given that the singular modes within each deformation-specific set are synthesized as weighted superpositions of the fiber’s native, mutually orthogonal eigenmodes, any output optical field can be effectively approximated by projecting it onto these bases, regardless of the deformation state from which they were derived. This representation remains valid as long as the full modal content is preserved. However, under sufficiently tight bending, the critical angle for total internal reflection may be exceeded, leading to radiative loss of higher-order modes into the cladding. In such cases, modal confinement deteriorates, and the resulting basis no longer spans the complete guided mode space of the undeformed fiber. Conversely, the set of singular modes derived from speckle fields recorded at increasingly tight bends exhibits a~progressive suppression of the highest-order guided modes, effectively narrowing the accessible modal bandwidth. Consequently, such a~basis can only fully represent speckle fields originating from equal or tighter bend configurations, as the loss of higher-order components precludes accurate reconstruction of fields corresponding to shallower bends. The evolution of these deformation-specific mode sets, together with their associated relative power distributions across linear deformation states, is visualized in~\cref{fig:modesSVD}. This trend is further quantified in ~\cref{fig:couplingStrengthsSVD} where the relative coupling strengths ($\widetilde{\sigma}_{\eta}^{(p)}$) are plotted as a~function of the singular mode index ($\eta$) for all sampled deformation states. The overlaid black dash-dotted and white dashed lines represent the average coupling profiles obtained from $\mathrm{LG}_{00}$ and $\mathrm{LG}_{\pm10}$ excitation, respectively, clearly demonstrating that orbital angular momentum (OAM)-carrying inputs promote stronger excitation of higher-order modes, thereby enriching the modal content accessible within the fiber’s transmission subspace. Notably, the near-complete overlap of the dashed mean-value curves corresponding to $\mathrm{LG}_{\pm10}$ inputs further reinforces the enhanced excitation. Additionally, a~clear and continuous decay in $\widetilde{\sigma}_{\eta}^{(p)}$ is observed as the fiber undergoes progressive deformation.  This trend is visually encoded via a~color gradient, ranging from $\widetilde{\sigma}_{\eta}^{(1)}$ (red) at $\Delta p_\text{mts} = \qty{0}{\milli\metre}$ to $\widetilde{\sigma}_{\eta}^{(N_{\mathrm{B}})}$ (purple) at $\Delta p_\text{mts} = \qty{50}{\milli\metre}$, thereby illustrating the effect of evolving bend configurations. Beyond the conventional representation of coupling strength, \cref{fig:couplingStrengthsSVD} also includes the cumulative coupling ratio $\bigl(\chi_{\eta}^{(p)}\bigr)$ and the associated error $\bigl(\varepsilon_{\eta}^{(p)}\bigr)$, both derived from the cumulative sum defined as $\zeta^{(p)}_{\eta} = \sum_{i=1}^{\eta}\sigma_{i}^{(p)}\bigl(\sum_{i=1}^{N_\mathrm{S}}\sigma^{(p)}_{i}\bigr)^{-1}$, which quantifies the progressive accumulation of modal power across increasing singular mode indices. Despite the fiber supporting $\approx{23}$ modes at the operating wavelength, coupling into the highest-order modes remains minimal, with $\widetilde{\sigma}_{N_{\mathrm{M}}}^{(p)}$ typically in the $\qtyrange[range-units=brackets,range-phrase=\text{--}]{50}{55}{\decibel}$ range. The first 18 singular modes already account for more than $\bar{\zeta}^{(p)}_{18}=\qty{99.99}{\percent}$ of the total transmitted power, indicating that the energy is concentrated within a~reduced modal subset.  This behavior results from the spot-scanning excitation scheme, which is effective for imaging applications through multimode fibers~\cite{Du:2024,Uhlirova:2024,Leite:2021,Tuckova:2021}, but does not provide uniform excitation or strict orthogonality the full modal bandwidth. Nevertheless, this approach facilitates a~systematic analysis of deformation-induced field evolution by associating the resulting speckle patterns with their respective excitation points on a~regular Cartesian grid across the fiber facet, thereby enabling interpretable spatial correlations between  between speckle fields, as illustrated in~\cref{fig:innerProductsBend_All}.


\begin{figure*}[t!]%
  \subcaptionbox{\label{fig:sparseCommonSVD_Modes}$\widetilde{\Psi}_{\rho}^{[i]}\leftarrow\mathrm{LG}_{00}$}{\includegraphics[width=0.476696476\textwidth]{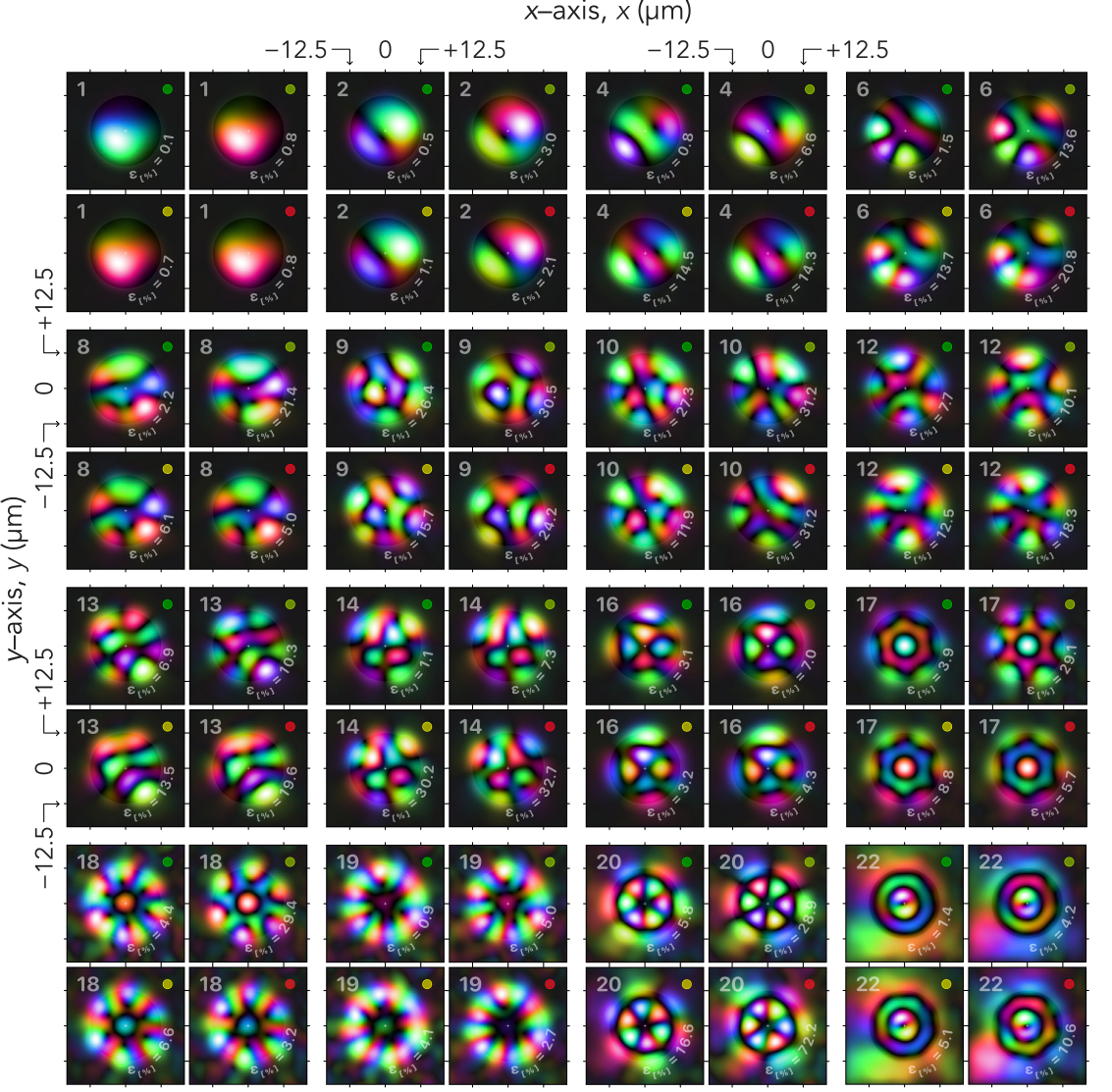}}\hfill%
  \subcaptionbox{\label{fig:sparseCommonSVD_InnerProducts}$\braket{\widehat{\Psi}_{\rho}}{\widetilde{\Psi}_{\rho^{\prime}}^{[i]}}\leftarrow\mathrm{LG}_{00}$}{\includegraphics[width=0.523303524\textwidth]{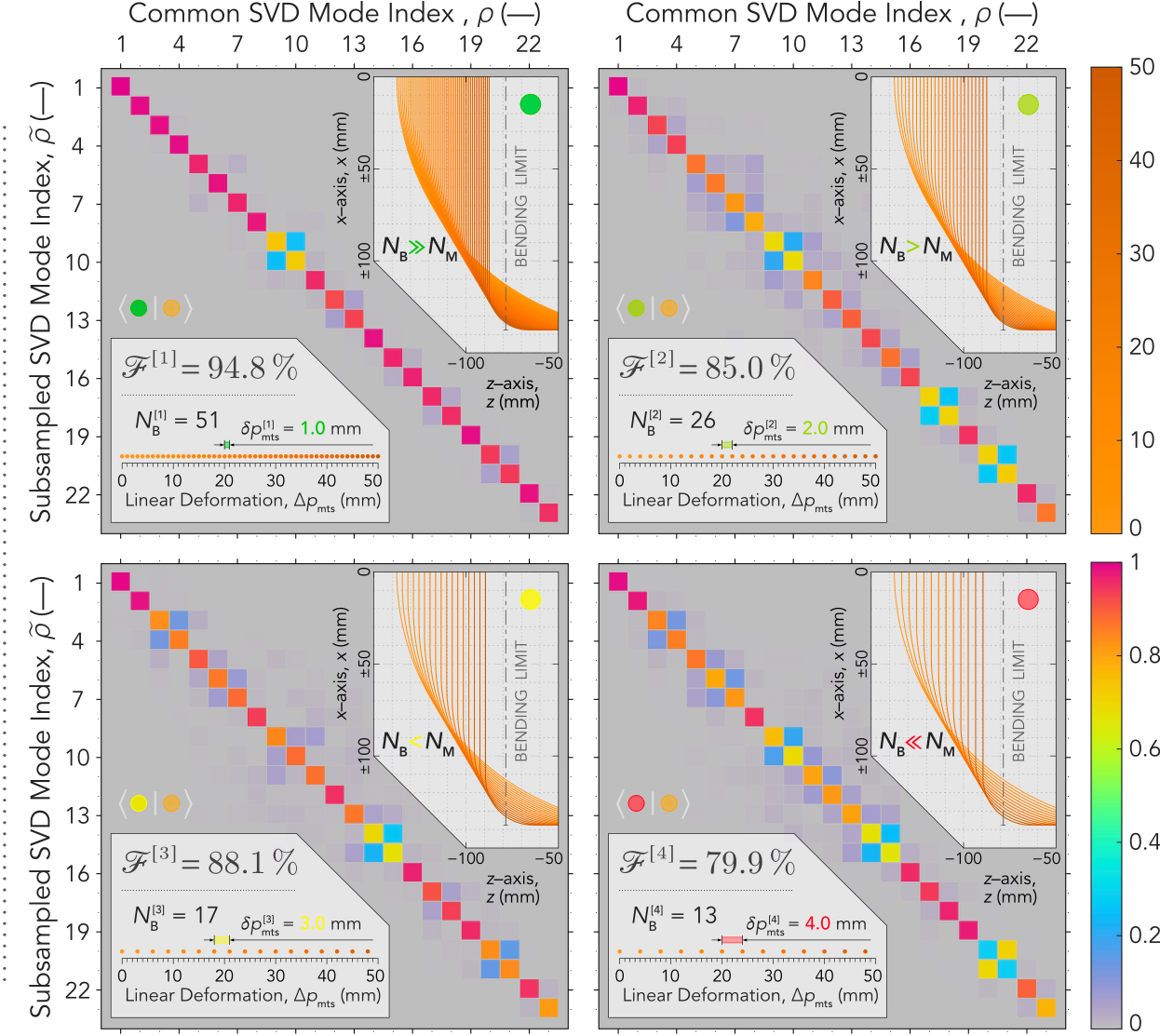}}%
	\caption{\label{fig:sparseCommonSVD_all}Stability analysis of the deformation-resolved global basis constructed from subsampled linear deformation states, benchmarked against the reference basis derived from the complete experimental deformation set. (a)~ Modal shape deviations are quantified for each subsampled configuration in terms of the associated per-mode reconstruction error ($\varepsilon_{[\unit{\percent}]}$), serving as a measure of individual mode infidelity. (b)~Overlap matrices evaluate the structural alignment between the full and subsampled common bases, generated using progressively coarser deformation sampling steps ($\delta{p}_{\text{mts}}^{[i]}$), resulting in reduced sets of deformation states ($\widetilde{N}_{\mathrm{B}}^{[i]}$), according to $\big\{\Delta\widetilde{p}_{\text{mts}}^{[i]}\big\}_{i=1}^{4} = \bigl\{\delta{p}_{\text{mts}}^{[i]}\cdot(p - 1)\mid\;\mathrel{{p}\in\mathbb{N}}\mathbin{,}\allowbreak\mathrel{{p}\leq\widetilde{N}_\mathrm{B}^{[i]}}\bigr\}\,\si{\milli\metre}
$, where $\big\{(\delta{p}_{\text{mts}}^{[i]},\widetilde{N}_{\mathrm{B}}^{[i]})\big\}_{i=1}^{4}
\mathbin{=}\allowbreak\big\{(\qty{1.0}{\milli\metre},51)\mathbin{,}\allowbreak(\qty{2.0}{\milli\metre},26)\mathbin{,}\allowbreak(\qty{3.0}{\milli\metre},17)\mathbin{,}\allowbreak(\qty{4.0}{\milli\metre},13)\big\}$. The preservation of modal consistency is quantitatively assessed using fidelity metrics, given by $\mathscr{F}^{[i]} = \text{tr}\bigl(N_{\mathrm{M}}^{-1}\bigl|\braket{\widehat{\Psi}_{\rho}}{\widetilde{\Psi}_{\rho^{\prime}}^{[i]}}\bigr|^2\bigr)\unit{\percent}$.}%
\end{figure*}%


\begin{figure*}[t!]%
  	\subcaptionbox{\label{fig:subset_1_CommonSVD_InnerProducts}$\Delta{\breve{p}}_{\text{mts}}^{\,[1]}$}{\includegraphics[width=0.25044685\textwidth]{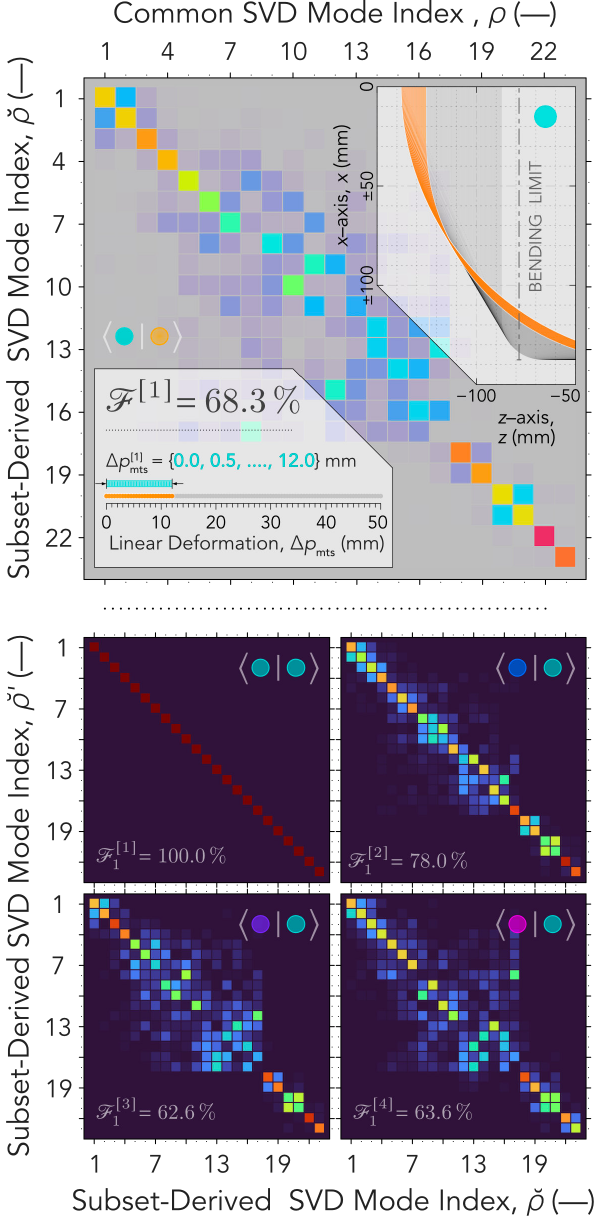}}\hfill%
  	\subcaptionbox{\label{fig:subset_2_CommonSVD_InnerProducts}$\Delta{\breve{p}}_{\text{mts}}^{\,[2]}$}{\includegraphics[width=0.23793502\textwidth]{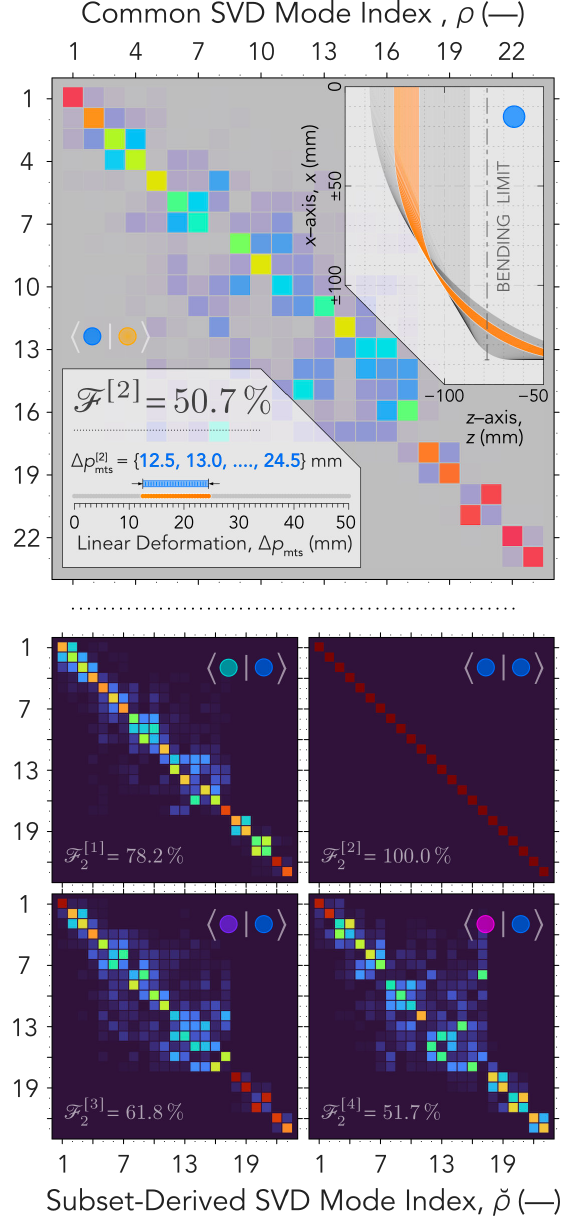}}\hfill%
    \subcaptionbox{\label{fig:subset_3_CommonSVD_InnerProducts}$\Delta{\breve{p}}_{\text{mts}}^{\,[3]}$}{\includegraphics[width=0.23793502\textwidth]{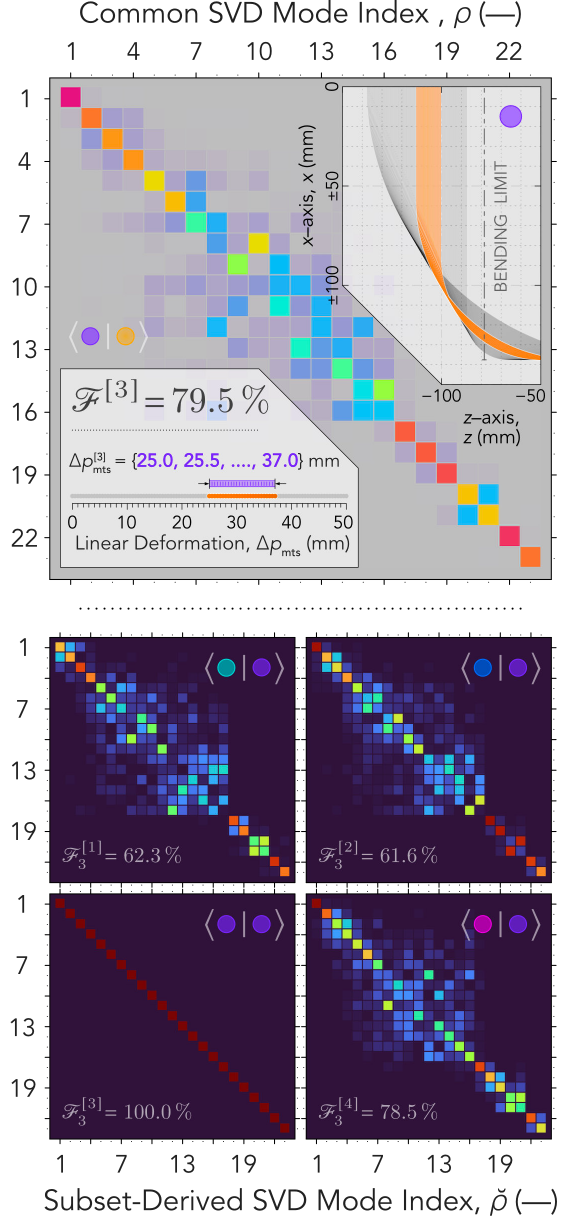}}\hfill%
    \subcaptionbox{\label{fig:subset_4_CommonSVD_InnerProducts}$\Delta{\breve{p}}_{\text{mts}}^{\,[4]}$}{\includegraphics[width=0.27368310\textwidth]{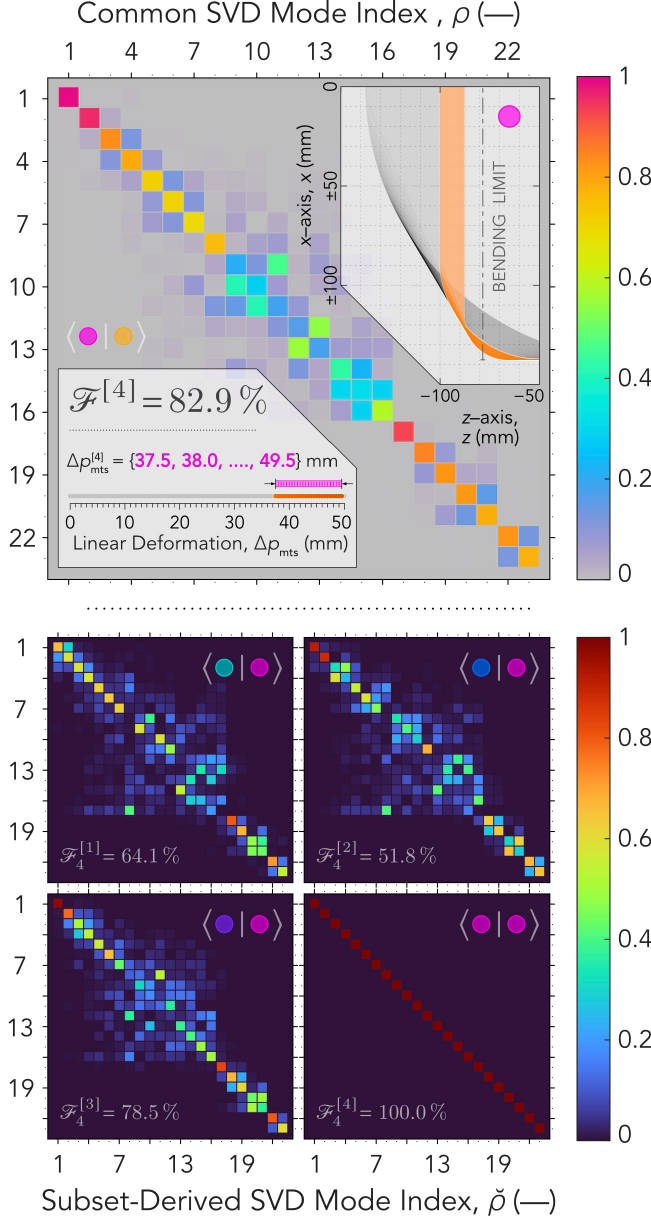}}%
	\caption{\label{fig:subsetCommonSVD_all}To assess the similarity of the bend-resolved basis across distinct ranges of mechanical deformation, overlap integrals, defined as $\braket{\widehat{\Psi}_{\rho}}{\breve{\Psi}_{\rho^{\prime}}^{[i]}}$, are computed between the globally constructed modal set and the subset-derived counterparts, as shown at the top of panels (a--d). The bottom of each panel presents the corresponding basis transformation matrices between the subset-derived SVD sets, defined as $\braket{\breve{\Psi}_{\rho}^{[i]}}{\breve{\Psi}_{\rho^{\prime}}^{[i^{\prime}]}}$.  In all cases, intermodal coupling under $\mathrm{LG}_{00}$ excitation remains predominantly confined to distinct groups of modes, ordered according to their singular values. The deformation-disjoint subsets considered in this analysis are defined as $\big\{\Delta\breve{p}_{\text{mts}}^{[i]}\big\}_{i=1}^{4} = \bigl\{ 0.5 \cdot (p - 1) + \gamma_{\mathrm{offset}}^{[i]} \mid\;\mathrel{p \in \mathbb{N}}\mathbin{,}\allowbreak\mathrel{p \leq \breve{N}_\mathrm{B}} \bigr\}~\si{\milli\metre}$, where $\big\{ \gamma_{\mathrm{offset}}^{[i]} \big\}_{i=1}^{4} = \qtylist[list-open-bracket=\{,list-close-bracket=\},list-units=brackets,list-separator={,},list-final-separator={\text{,}\,}]{0.0;12.5;25.0;37.5}{\milli\metre}$ and $\breve{N}_\mathrm{B}=25$.}%
\end{figure*}%


As expected, the singular modes within each set associated with a~specific deformation exhibit perfect orthogonality, formally expressed as $\braket{\Psi_{\eta}^{(p)}}{\Psi_{\eta^{\prime}}^{(p^{\prime})}} = \delta_{pp^{\prime}}\delta_{\eta\eta^{\prime}}$. However, this orthogonality does not generally extend across deformation states, where the inner product between modes is typically nonzero, i.e. $\braket{\Psi_{\eta}^{(p)}}{\Psi_{\eta}^{(p^{\prime})}}\neq{0}$ for ${p\neq{p}^{\prime}}$. The correlation matrices, together with axial cross-sections of selected singular modes for $\eta=\{4,7\}$  under varying $\Delta{p}_\text{mts}$, are presented in~\cref{fig:innerProductSVD}. These matrices exhibit a dynamic structure similar to that of the original correlation matrices derived directly from the speckle patterns, shown in~\cref{fig:innerProductsBend_All}. Within this framework, each per-deformation basis captures the dominant deformation-induced variations in the output fields and, under ideal excitation and detection conditions, spans a~space isomorphic to the dimensionality of the fiber’s guided mode space. The field cross-sections taken along the Cartesian axes at $(x_0, y_0) = \bigl\{(\pm{r}, 0), (0,\pm{r})\bigr\}$ in \cref{fig:innerProductSVD_4_LG00,fig:innerProductSVD_7_LG00} demonstrate mutual orthogonality. Notably, the correlation matrices and associated fields differ between excitations with $\mathrm{LG}_{00}$ and $\mathrm{LG}_{\pm10}$ modes, while excitations using OAM-carrying modes of opposite $\ell$ sign but identical $\eta$ yield nearly identical cross-sectional profiles, as shown in~\cref{fig:innerProductSVD_4_LG-10,fig:innerProductSVD_7_LG+10}, indicating near-orthogonality under symmetric angular excitation.


To further reduce the dimensionality of the deformation-dependent speckle field representation while retaining an orthonormal basis aligned with the physically relevant modal subspace of the fiber, we introduce a second-stage SVD. This global decomposition is applied to the concatenated set of singular modes derived independently for each deformation state, yielding a~deformation-resolved global basis that captures the dominant structural variations induced by mechanical perturbations. To accomplish this, we concatenate all singular mode sets corresponding to various bend deformations into a unified, overcomplete set $\Psi_\varsigma\!\left(r,\theta\right)$, in which the original indices $\eta$ and $p$ are remapped to a~single index $\varsigma = \varsigma(\eta,p)\in\mathbb{N}$, resulting in a~loss of orthogonality between singular modes associated with different deformation $p$ states. According to~\cref{eq:inner_product}, one can construct a~unified correlation super-matrix defined by $\braket{\Psi_{\varsigma}}{\Psi_{\varsigma^{\prime}}}=\widehat{M}_{\varsigma\varsigma^{\prime}}$, where $\widehat{\mat{M}} \in \mathbb{C}^{(\mathcal{r}_\mathrm{X}N_\mathrm{B}) \times (\mathcal{r}_\mathrm{X}N_\mathrm{B})}$. Applying the factorization outlined in~\cref{eq:svd}, followed by the construction procedure described in~\cref{eq:construction_svd}, leads to the formation of a~robust modal basis
\begin{equation}\label{eq:robust_svd}
	\widehat{\Psi}_{\rho}\!\left(r,\theta \right) = \sigma_{\!\rho}^{-\frac{1}{2}}\sum_{\varsigma=1}^{\mathcal{r}_\mathrm{X}N_\mathrm{B}}V_{\varsigma\rho}^{\ast}\,\Psi_{\varsigma}(r,\theta),
\end{equation}
which defines a~common, globally orthonormal set of modes that efficiently spans the fiber’s deformation-relevant modal subspace, while capturing the dominant structural variations induced by bending. This basis reflects the underlying physical behavior of the system, rather than merely ensuring formal completeness across the full modal space. It is crucial to emphasize that a~universal, deformation-resolved modal basis cannot be obtained through a~direct SVD of a~correlation matrix constructed from the aggregate ensemble of speckle fields across all input displacements and bending configurations. Such an approach neglects the structured mapping between deformation states and their corresponding modal subspaces, thereby failing to preserve the deformation-specific transformations required for constructing a~coherent, physically meaningful global basis. This common deformation-related basis set, shown in~\cref{fig:robustSVD}, exhibits enhanced spatial symmetry compared to the individual per-deformation bases depicted in~\cref{fig:modesSVD}. Notably, the highest-order modes $\widehat{\Psi}_{\rho}\mid{\rho\in\{\num{20},\dots,\num{23}\}}$ closely resemble the fiber’s doubly degenerate, propagation-invariant odd and even $\mathrm{LP}_{\ell n}\mid(\ell,n)\in\big\{(3,2),(1,3)\big\}$ modes under the weakly guiding approximation, as illustrated in~\cref{fig:approximatedLPmodes}, despite the limited excitation coverage provided by the spot-like input basis. In contrast, other singular modes, such as $\widehat{\Psi}_{\rho}\mid{\rho\in\{\num{17},\num{18},\num{19}\}}$, resemble axially power-balanced linear combinations of the fiber’s $\mathrm{LP}_{\ell n}\mid(\ell, n)\in\big\{(0,3),(5,1)\big\}$ modes, characterized by nearly indistinguishable propagation constants, as illustrated in the middle plot of~\cref{fig:matricesLPmodes}.%


We observed that the spatial structure of the global basis remains consistent regardless of the subset of speckle patterns selected from the input basis, provided that the condition $N_{\mathrm{S}} \geq N_{\mathrm{M}}$ is strictly satisfied. Moreover, selecting a~subset of speckles from an off-center, laterally shifted excitation grid yields different singular mode profiles in the first-stage decomposition, yet the spatial configuration of the global basis remains unchanged, apart from a~global phase factor. This invariance also holds when the number of deformation states used in the second-stage factorization is reduced, as shown in~\cref{fig:sparseCommonSVD_all}. In this context, the spatial structure of the global modes remains robust, provided the sampled deformation space spans a sufficient dimensionality to capture the full modal content of the fiber. Theoretically, it is essential to include a number of distinct deformations at least comparable to the fiber’s modal bandwidth; otherwise, the dominant intermodal coupling effects introduced by bending cannot be fully represented. Nevertheless, as demonstrated in~\cref{fig:sparseCommonSVD_InnerProducts}, subsampling the set of deformation states according to $\big\{\Delta\widetilde{p}_{\text{mts}}^{[i]}\big\}_{i=1}^{4} = \bigl\{\delta{p}_{\text{mts}}^{[i]}\cdot(p - 1)\mid\;\mathrel{{p}\in\mathbb{N}}\mathbin{,}\allowbreak\mathrel{{p}\leq\widetilde{N}_\mathrm{B}^{[i]}}\bigr\}\,\si{\milli\metre}
$, where $\big\{(\delta{p}_{\text{mts}}^{[i]},\widetilde{N}_{\mathrm{B}}^{[i]})\big\}_{i=1}^{4}
\mathbin{=}\allowbreak\big\{(\qty{1.0}{\milli\metre},51)\mathbin{,}\allowbreak(\qty{2.0}{\milli\metre},26)\mathbin{,}\allowbreak(\qty{3.0}{\milli\metre},17)\mathbin{,}\allowbreak(\qty{4.0}{\milli\metre},13)\big\}$ and evaluating the overlap integral $\braket{\widehat{\Psi}_{\rho}}{\widetilde{\Psi}_{\rho^{\prime}}^{[i]}}$ between the fully spanned basis (shown in \cref{fig:robustSVD}) and its subsampled counterparts (depicted in \cref{fig:sparseCommonSVD_Modes}), demonstrates that the spatial consistency of the bases is preserved. Incorporating fidelity metrics, defined as $\mathscr{F}^{[i]} = \text{tr}\bigl(N_{\mathrm{M}}^{-1}\bigl|\braket{\widehat{\Psi}_{\rho}}{\widetilde{\Psi}_{\rho^{\prime}}^{[i]}}\bigr|^2\bigr)\si{\percent}$ facilitates a~precise and rigorous quantitative assessment of mode alignment across variations induced by sparse subsampling. For scenarios where $\widetilde{N}_{\mathrm{B}} \geq N_{\mathrm{M}}$, the fidelity reaches $\mathscr{F}^{[1]} = \SI{94.8}{\percent}$ and $\mathscr{F}^{[2]} = \SI{85.0}{\percent}$ for step sizes $\delta p_{\text{mts}}^{[1]} = \qty{1.0}{\milli\metre}$ and $\delta p_{\text{mts}}^{[2]} = \qty{2.0}{\milli\metre}$, respectively. Notably, the sparse overlap matrix retains a~predominantly diagonal structure even for subsampled deformation sets that do not fully span the fiber’s modal space. A step size of $\delta p_{\text{mts}}^{[3]} = \qty{3.0}{\milli\metre}$ yields a~fidelity of $\mathscr{F}^{[3]} = \SI{88.1}{\percent}$, while $\delta p_{\text{mts}}^{[4]} = \qty{4.0}{\milli\metre}$ yields $\mathscr{F}^{[4]} = \SI{79.9}{\percent}$. Interestingly, the fidelity associated with $\widetilde{N}_\mathrm{B}^{[3]} = \num{17}$ surpasses that for $\widetilde{N}_\mathrm{B}^{[2]} = \num{26}$. However, in the second case where $\widetilde{N}_\mathrm{B} > N_\mathrm{M}$, power coupling appears to be confined primarily to neighboring mode pairs (degenerate pairs), as shown in~\cref{fig:sparseCommonSVD_InnerProducts}. This behavior is likely driven by linear combinations of modes sharing equivalent modal GD, characterized by nearly identical propagation constants (see attachment~\cref{fig:approximatedLPmodes_all}). For example, compare the subsampled singular modes $\big\{ \widetilde{\Psi}_{\rho}^{[i]}\mid\rho\in\{9, 10\},\,i\in\{1,2\}\big\}$,  as depicted in~\cref{fig:sparseCommonSVD_Modes}, or the overlap integral matrices for subsampled singular modes $\big\{ \widetilde{\Psi}_{\rho}^{[i]}\mid\rho\in\{21, 22\},\,i\in\{2,4\}\big\}$, which closely resemble the $\mathrm{LP}_{32}$ mode.%


As shown in the cross-section of the average cumulative power (\cref{fig:powerThroughFiber_Field}) and the orthogonal field cross-sections of singular modes with quasi-equivalent coupling strength (\cref{fig:innerProductSVD_4_LG00,fig:innerProductSVD_7_LG00}), the complexity of modal interactions increases with deformation, as greater linear displacement induces higher curvature in the bent fiber segments. To better isolate the contributions from specific deformation intervals, a~reduced set of common bases was constructed by partitioning the full deformation range into four disjoint subsets, according to $\big\{\Delta\breve{p}_{\text{mts}}^{[i]}\big\}_{i=1}^{4} = \bigl\{ 0.5 \cdot (p - 1) + \gamma_{\mathrm{offset}}^{[i]} \mid\;\mathrel{p \in \mathbb{N}}\mathbin{,}\allowbreak\mathrel{p \leq \breve{N}_\mathrm{B}} \bigr\}~\si{\milli\metre}$, where $\big\{ \gamma_{\mathrm{offset}}^{[i]} \big\}_{i=1}^{4} = \qtylist[list-open-bracket=\{,list-close-bracket=\},list-units=brackets,list-separator={,},list-final-separator={\text{,}\,}]{0.0;12.5;25.0;37.5}{\milli\metre}$ and $\breve{N}_\mathrm{B}=25$. The overlap integral $\braket{\widehat{\Psi}_{\rho}}{\breve{\Psi}_{\rho^{\prime}}^{[i]}}$ serves to evaluate the contributions of disjoint deformation subsets by comparing the fully spanned, bend-resolved singular basis shown in~\cref{fig:robustSVD} to its subset-derived counterparts, thereby elucidating shared spatial modal characteristics, as illustrated at the top of~\cref{fig:subsetCommonSVD_all}. As shown at the top of~\cref{fig:subsetCommonSVD_all}, the last two subset-derived mode sets, corresponding to the deformation ranges $\bigl\{\Delta{\breve{p}}_{\text{mts}}^{\,[i]}\bigr\}_{i\in\{3,4\}}\mathbin{=}\allowbreak\bigl\{ \{{25.0,25.5,\dots,37.0}\}\mathbin{,}\allowbreak\{{37.5,38.0,\dots,49.5}\}\bigr\}\,\unit{\milli\metre}$, originate from narrower deformation intervals and therefore capture more progressive variations in the output field dynamics. These subsets achieve fidelities of $\mathscr{F}^{[3]} = \SI{79.5}{\percent}$ and $\mathscr{F}^{[4]} = \SI{82.9}{\percent}$, respectively, indicating strong alignment with the global modal framework.  In contrast, the deformation sets associated with lower curvature, characterized by less progressive field variations (see \cref{fig:innerProductSVD_4_LG00,fig:innerProductSVD_7_LG00})and defined as $\bigl\{\Delta{\breve{p}}_{\text{mts}}^{\,[i]}\bigr\}_{i\in\{1,2\}}\mathbin{=}\allowbreak\bigl\{ \{{0.0,0.5,\dots,12.0}\}\mathbin{,}\allowbreak\{{12.5,13.0,\dots,24.5}\}\bigr\}\,\unit{\milli\metre}$, exhibit comparatively lower fidelity with respect to the fully spanned common basis, achieving $\mathscr{F}^{[1]} = \SI{68.3}{\percent}$ and $\mathscr{F}^{[2]} = \SI{50.7}{\percent}$, reflecting weaker alignment with the global modal structure. The lower fidelity of the second deformation subset is attributed to an SVD ambiguity, wherein the modes $\breve{\Psi}_{\rho}^{[2]}\mid{\rho\in\left\{\num{21},\num{22}\right\}}$ are interchanged. Resolving this ambiguity would increase the alignment with the fully spanned basis to approximately $\approx\qty{70.0}{\percent}$, indicating a more accurate correspondence between subsets and the global modal set. The transformation matrices between disjoint subsets, shown at the bottom of~\cref{fig:subsetCommonSVD_all},  exhibit the strongest correlations between adjacent deformation ranges, and this trend persists even when the deformation span is divided into smaller subsets ($N_\mathrm{M} > \breve{N}_\mathrm{B}$). As shown in~\cref{fig:subsetCommonSVD_all}, the lower-order modes $\breve{\Psi}_{\rho}\mid{\rho\in\{\num{1},\dots,\num{7}\}}$, consistently display a~predominantly diagonal structure, which becomes more pronounced for subsets corresponding to tighter bends. These modes, strongly confined to the fiber core, are less prone to intermodal coupling and thus form a~stable component of the modal space. In contrast, middle-order modes, less confined and with effective indices ($n_\mathrm{eff}$) approaching that of the cladding, are more vulnerable to distortion and coupling. These modes $\breve{\Psi}_{\rho}\mid{\rho\in\{\num{8},\dots,\num{16}\}}$ often become ambiguous as they blend with one another, showing variable coupling patterns due to their heightened sensitivity to bending or deformation. Conversely, the highest-order modes $\breve{\Psi}_{\rho}\mid{\rho\in\{\num{17},\dots,\num{23}\}}$ are effectively isolated, with minimal power exchange. As these modes approach their cutoff under tight bends, where the critical angle $\theta_\mathrm{c}$ is exceeded, they transition into radiative or leaky states. This  attenuation shortens their effective interaction length, thereby limiting the possibility of meaningful intermodal coupling. Even in the presence of spatial overlap, the combination of high loss and limited propagation distance renders energy exchange between these modes negligible.%

\section{Discussion}%

\noindent Fiber deformation in practical experimental settings rarely conforms to an idealized planar geometry, as residual torsion, intrinsic twisting, and microbending irregularities perturb the spatial refractive index distribution beyond the monotonic transverse refractive index gradients associated with planar bending, fundamentally altering intermodal coupling mechanisms and complicating the interpretation of modal dynamics. To systematically isolate and examine curvature-induced modal behavior under reproducible and geometrically constrained conditions, the fiber was mounted in a~pre-bent U-shaped configuration and actuated by a~linear MTS driving a~V-groove adapter, as illustrated in ~\cref{fig:volumetricSpeckle_Fiber,fig:optical_setup}. This arrangement produced two symmetric bends with systematically decreasing radii in a well-defined plane, while progressively forming intermediate and lateral straight segments between the curved regions and the permanently straight sections near the fiber’s proximal and distal facets, as detailed in~\cref{sec:bending_geometry}. The resulting geometry imposed deformation-induced intermodal coupling from the initial state, suppressing trivial phase accumulation associated with uncoupled propagation in straight fibers and ensuring that all recorded speckle patterns originated from structurally meaningful modal interactions. Moreover, its bilateral symmetry and geometric regularity facilitate rigorous FD-BPM modeling, enabling reproducible and physically consistent simulations of modal evolution under prescribed curvature profiles and segment lengths, as detailed in~\cref{sec:transmission_throught_fiber}.


Near-field speckle patterns from excitation proximate to the core--cladding interface exhibited markedly reduced variability under deformation, indicating inefficient re-coupling of higher-order, weakly confined modes into the strongly core-guided modal subspace. Instead, power associated with these modes likely leaked into the cladding, contributing negligibly to deformation-resolved dynamics. This behavior aligns with prior imaging-based studies demonstrating that fiber deformation predominantly perturbs core-confined modes, leading to pronounced speckle scrambling, while weakly confined peripheral modes undergo gradual attenuation and spatial redistribution governed by the bend radius and axial extent of deformation~\cite{Mekhail:2024,Boonzajer:2018,Li:2021:b}. Dynamic transformations in the speckle field, driven by lateral input displacements, were further modulated by OAM. Although the continuous evolution of these patterns provided insight into intermodal coupling processes, it led to an overestimation of the fiber’s effective modal bandwidth. Dimensionality reduction via SVD was employed to compress the diverse speckle fields associated with each fiber deformation state into reduced orthonormal bases spanning the accessible modal space. A~subsequent second-stage SVD, applied to the concatenated per-deformation bases, yielded a compact bend-resolved modal basis that robustly encapsulates the principal conformational variations induced by fiber deformation~\cite{Xu:2023}. Despite coupling variations under OAM excitation, the second-stage SVD consistently converges to a~stable modal basis exhibiting highly reproducible spatial profiles across all datasets. The extracted basis remains structurally robust across subsets of speckle fields or reduced translation-stage sampling ranges. Residual variations were primarily confined to global phase factors, attributable to the intrinsic sign ambiguity of the SVD, and to column permutations caused by near-degenerate singular values.


In the extracted common basis, modes associated with the lowest and highest coupling strengths exhibit pronounced spatial symmetry and closely resemble the degenerate eigenmodes of an unperturbed step-index fiber. This basis constitutes an optimal linear combination of guided modes that efficiently captures the dominant deformation-induced power exchange across all configurations within the examined U-shaped geometry~\cite{Lan:2020}. Its structure is governed by the applied mechanical deformation and remains consistent under both conventional and OAM-carrying excitation. Modal sensitivity to external perturbations varies across the singular spectrum, with higher-order modes exhibiting increased sensitivity suitable for real-time detection of mechanical impairments, while lower-order modes maintain structural stability preferable for MDM~\cite{Qiao:2024}. Optimized transmission strategies can exploit this structure by selecting modes with stable differential GD or by restricting excitation to robust modal subsets, thereby enabling efficient deployment of FMFs that combine the high-capacity benefits of multimode systems with the propagation stability and signal integrity of single-mode links~\cite{Cifuentes:2025,Sillard:2022}.


Although the proposed method exhibits strong performance in resolving deformation-dependent modal structures, its accuracy is subject to limitations inherent to experimental implementation. Artifacts arising from cladding-coupled light, ambient fluctuations, and structured background speckle may affect the reliability of the SVD, particularly for modes with low excitation efficiency. In contrast to numerical simulations with absorbing boundary conditions and controlled noise levels, experimental measurements may contain persistent cladding-guided contributions resulting from light leakage or unintended coupling into the cladding. These factors can introduce spurious components into the modal basis, especially when the background signal approaches or exceeds the energy of the weakest guided modes, potentially leading to the emergence of nonphysical modes that may be misidentified, along with perturbations in the ordering of physically meaningful singular vectors. Applying SVD directly to a~correlation matrix constructed from the full ensemble of speckle fields, which includes all input displacements and deformation states, may obscure the underlying structure associated with specific mechanical configurations. This approach blends deformation-specific modal variations, thereby reducing the interpretability and physical relevance of the basis. In contrast, a~two-stage SVD applied to deformation-specific modal sets preserves this structure and enables the construction of a~compact global basis that captures the dominant variations across all fiber conformations.


Adapting the proposed two-stage SVD framework to the C-band introduces practical and physical challenges despite its inherent wavelength agnosticism. The reduced number of spatial modes supported aligns naturally with the few-mode regime targeted in modern MDM systems, including coherent architectures capable of resolving intermodal phase for digital equalization, albeit with limited modal diversity~\cite{Fontaine:2012,Fazea:2021}. Increased differential GD and tighter transverse confinement at C-band wavelengths reduce alignment tolerance and increase susceptibility to calibration errors, potentially compromising the accuracy of modal fidelity measurements in long-haul or high-speed links essential for reliable modal decomposition. Together with wavelength-dependent variations in coupling dynamics and spatial mode distributions, these effects degrade the experimental ability to discriminate individual modes, increasing intermodal crosstalk and ambiguity in the reconstructed modal set, and necessitating re-optimization of both excitation and detection strategies. These constraints not only limit the achievable mode separation in MDM systems but also undermine speckle-based modal reconstruction techniques that rely on stable, high-contrast interference patterns for precise modal discrimination. Detection in this band relies on InGaAs sensors, which have lower pixel density and dynamic range, making photonic integrated circuit platforms a~more scalable and robust alternative~\cite{Lu:2024,Butow:2024}. Despite these constraints, the framework remains directly transferable to C-band implementations.

\section{Conclusion}%

\noindent We have presented a~bend-resolved modal decomposition framework that constructs a~compact basis directly from experimentally measured speckle fields across controlled mechanical deformations. Rather than approximating fiber eigenmodes, the resulting basis captures the dominant degrees of freedom that span the space of deformation-induced output field variations. Each bend-resolved basis reflects the statistical structure of speckle patterns associated with a~specific range of mechanical states. As the deformation span changes, new modal subspaces can emerge, demonstrating the basis’s sensitivity to mechanical perturbations and its ability to encode deformation-dependent modal transformations. This approach is both model-free and data-driven, requiring no prior knowledge of the fiber’s geometry or wave equation solutions, relying solely on a~representative set of deformation-induced speckle fields. The two-stage SVD procedure first extracts locally orthonormal modes for each deformation state, then performs a~global decomposition that preserves deformation-specific variation while yielding a~compact and globally consistent basis.


The proposed method can be integrated into fiber-based imaging and communication systems to track mechanical disturbances, reconstruct phase-scrambled signals, and improve system robustness under real-world conditions. In imaging, the approach enables compensation for modal scrambling in flexible endoscopes or dynamically deforming systems. In coherent mode-division multiplexing, it may support adaptive demultiplexing, enhance MIMO channel estimation, or detect tampering and motion beyond a~calibrated mechanical envelope. Although exact inversion of deformation from modal transitions is limited by degeneracy and non-uniqueness, the extracted bend-resolved modes provide a~practical and informative basis for monitoring and interpreting fiber behavior under mechanical stress. Further investigation into selective mode launching and the isolation of deformation-sensitive modes could support more direct mapping between spatial mode structure and fiber deformation, potentially enabling the inference of bending profiles along the perturbed fiber.

\begin{acknowledgments}%
The work was supported by the Horizon 2020 Future and Emerging Technologies Open grant agreement "Super-pixels" No.\,829116 and the EPSRC (grant No.~EPSRC EP/\-T009012/1).%
\end{acknowledgments}%
\section*{Data Availability Statement}
The data supporting the findings of this study are available from the University of Glasgow repository at \href{https://doi.org/10.5525/gla.researchdata.2038}{\texttt{doi.org/10.5525/gla.researchdata.2038}}.
\bibliographystyle{aipnum4-1}
\bibliography{aiptemplate}%
\appendix%

\section{\label{volumetric_visualization}Volumetric visualization}%


\noindent The volumetric visualization in \cref{fig:volumetricSpeckle_Fiber} presents a~selected set of experimentally acquired speckle patterns, generated under constant injection with a~fixed displacement at the proximal facet, such that  $E_{\mu}^{(p)}\!(r,\theta, z=0) \equiv E_{\mu}(r,\theta)$. These patterns systematically evolve as the fiber deformation is progressively altered by the MTS. Although the same spatial input position is used for all cases, the injected fields differ by employing distinct members of the set $\big\{\mathrm{LG}_{\ell0}\mid{\ell \in \{\pm 1, 0\}}\big\}$, which excite fiber modes with varying efficiencies and produce different output speckle fields. Despite these differences, the cross-section of the quarter-volume in~\cref{fig:volumetricSpeckle_Fiber} reveals substantial similarities among the output fields. The inclusion of OAM enhances the fraction of power coupled into higher-order modes that would otherwise couple into the fundamental mode when using a~Gaussian spot. 

While the speckle patterns vary due to differences in modal power distribution, the total cumulative power across all deformations  remains highly correlated across different LG mode injections (see \cref{fig:powerThroughFiber_Field}). This is also reflected in the cumulative intensity summed over all input fields at specific transverse slices $(x_0, y_0) = \bigl\{(\pm{r}, 0), (0,\pm{r})\bigr\}$. It is also evident that the output power consistently accumulates along the negative $x$-direction, with a~dominant symmetry about the $y$-axis. This behavior is not observed in simulations of an undeformed straight fiber unless all input fields on the Cartesian grid include an additional phase tilt. Specifically, when the central input field $(\upsilon,\nu)=(0,0)$ has negligible phase tilt relative to the fiber axis (that is, when the wavefront from the SLM’s first diffraction order is nearly parallel to the proximal facet), the output power forms a~spatially uniform, flat-top profile. Only when all displaced input fields are tilted in a~common direction does the output power shift accordingly. In the experimental setup, global tilt was minimized. Therefore, the observed power accumulation along the negative $x$-axis, with symmetry around the $y$-axis, is attributed to the U-shaped bend of the fiber. As previously noted (see \cref{fig:setup_bend}), the fiber initially bends from the negative to the positive $x$-direction within the $xz$-plane, while maintaining a constant transverse position in the $xy$-plane.


\section{\label{sec:bending_matrices}Correlative Analysis}%


\begin{figure*}[t!]%
  \subcaptionbox{\label{fig:bendMatrices_LG-10}$|\mathbf{C}_{\mu}|\leftarrow\mathrm{LG}_{-10}$}{\includegraphics[width=0.30686717\textwidth]{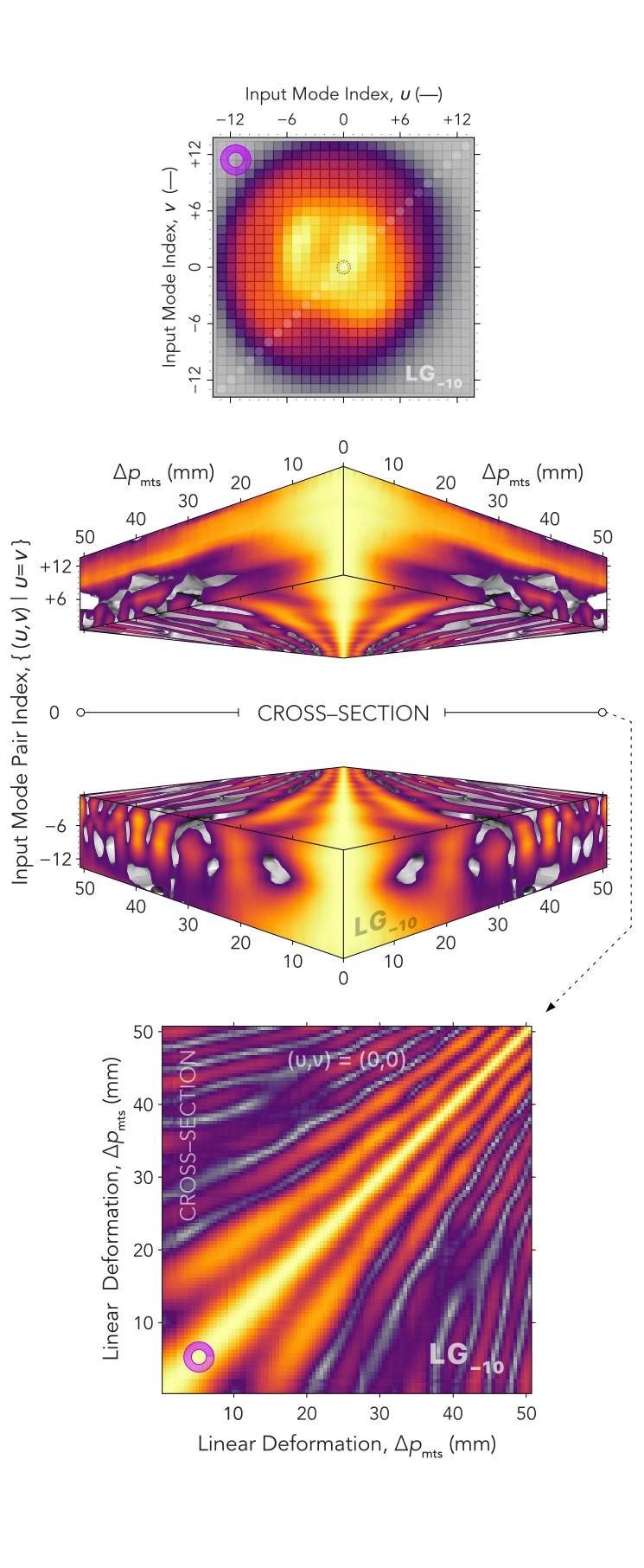}}\hfill%
  \subcaptionbox{\label{fig:bendMatrices_LG00}$|\mathbf{C}_{\mu}|\leftarrow\mathrm{LG}_{00}$}{\includegraphics[width=0.386265643\textwidth]{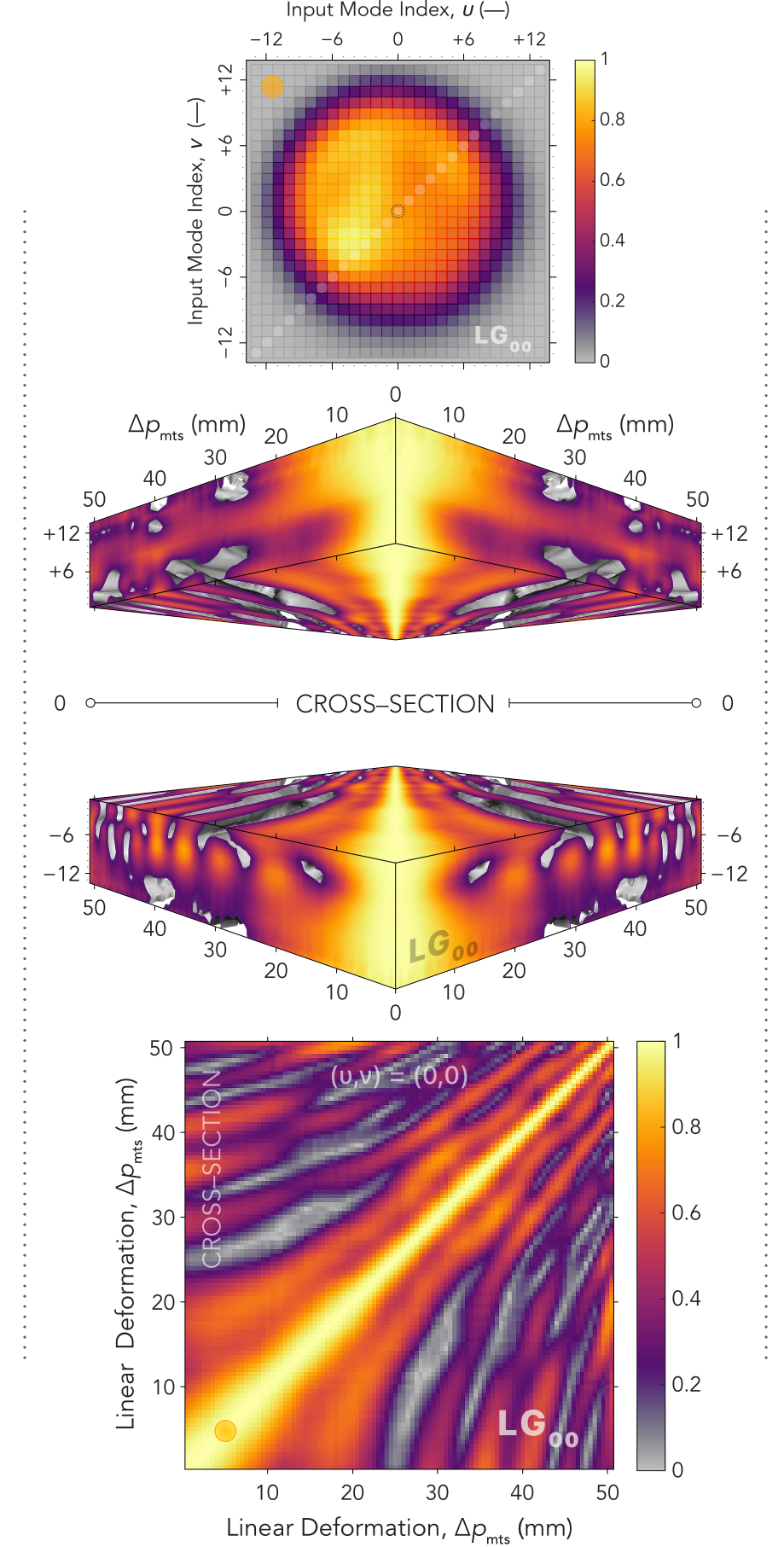}}\hfill%
  \subcaptionbox{\label{fig:bendMatrices_LG+10}$|\mathbf{C}_{\mu}|\leftarrow\mathrm{LG}_{+10}$}{\includegraphics[width=0.30686717\textwidth]{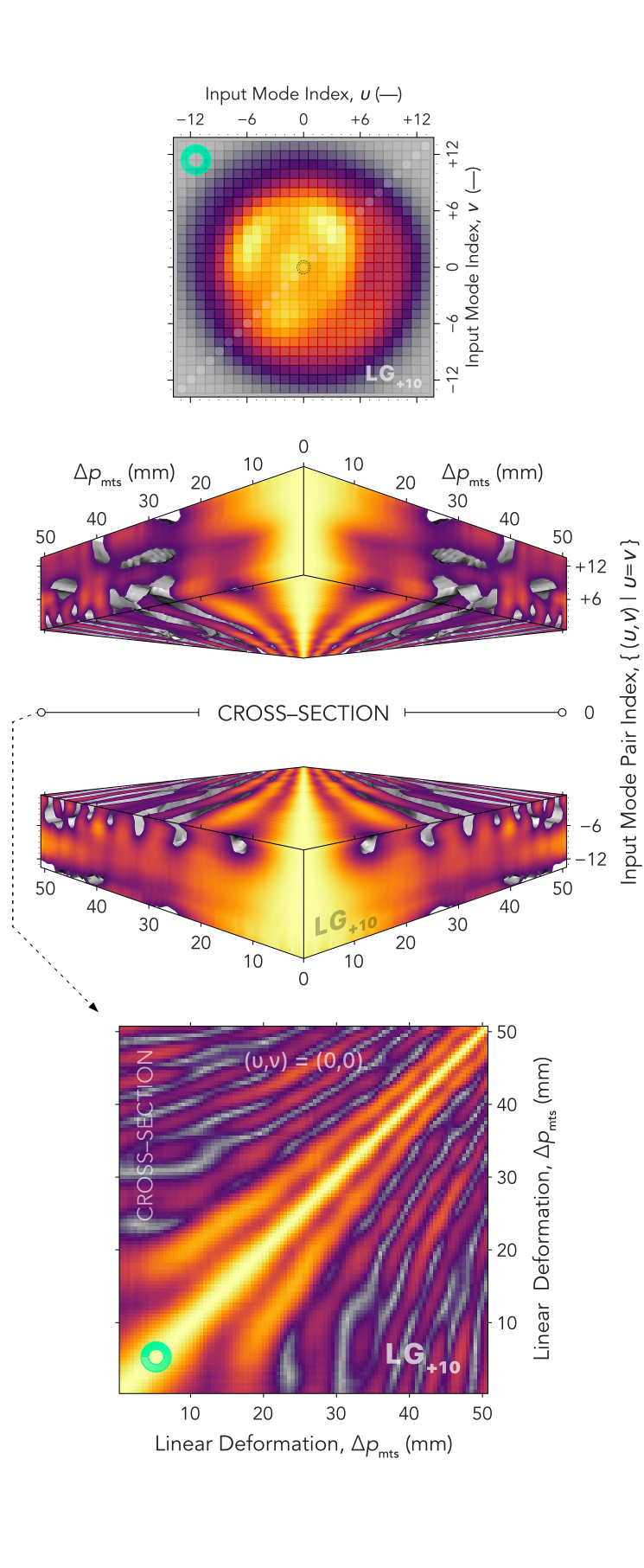}}%
	\caption{\label{fig:innerProductsBend_All}Correlative analysis of speckle patterns emerging in response to linear deformation under constant displacement of the injected beam for (a,\,c)~coupling with OAM-carrying $\mathrm{LG}_{\pm10}$ modes and (b)~coupling with the Gaussian $\mathrm{LG}_{00}$ mode. Top: Total power through all linear deformations as a~function of displacement of input field, $\overline{P}(\upsilon,\nu) = \mathrm{vec}^{-1}_{\upsilon\nu}\bigl\{(N_\mathrm{B}N_\mathrm{px})^{-1}\sum_{p=1}^{N_\mathrm{B}}\sum_{i=1}^{N_\mathrm{px}}\big|{S_{\mu}^{(p)}\!(x_i,y_i)}\big|^{2}\bigr\}$. Middle: Stacked visualization of correlation matrices constructed from speckle fields under varying deformation $\braket{\widehat{S}_{\mu}^{(p)}}{\widehat{S}_{\mu}^{(q)}} = [\mathbf{C}_{\mu}]_{pq}$, with fixed beam injection position for input mode pairs $\bigl\{(\upsilon,\nu)\mid\upsilon=\nu\neq{0}\bigr\}$. Bottom: Correlation matrix at $(\upsilon,\nu)=(0,0)$. Notes: $|\mathbf{C}_{\mu}|\in[0,1]$, where lower limit indicates completely uncorrelated fields and upper limit perfect correlation in amplitude and phase.}%
\end{figure*}%


\noindent To analyze the dynamic redistribution of the optical field under fiber deformation, correlation matrices of overlap integrals are presented in~\cref{fig:innerProductsBend_All}. These matrices represent the inner products between normalized speckle fields measured at different deformation states for a~fixed injection position with entries $[\mathbf{C}_{\mu}]_{pq} = \braket{\widehat{S}_{\mu}^{(p)}}{\widehat{S}_{\mu}^{(q)}}$, where $\widehat{S}={S}/\|{S}\|$ is normalized speckle, index $q = p^{\prime}$ and $\mat{C}_{\mu} \in \mathbb{C}^{N_\mathrm{B} \times N_\mathrm{B}}$. The central volumetric plot illustrates the evolution of these correlation matrices for input modes along the diagonal of the displacement grid $\bigl\{(\upsilon, \nu) \mid \upsilon = \nu\bigr\}$, as indicated in the top panel of~\cref{fig:innerProductsBend_All}. The highlighted cross-section in the volumetric visualization indicates the position of the correlation matrix for $\mu(0,0)$, also displayed separately at the bottom. While the cumulative transmitted power for $\mathrm{LG}_{00}$ is nearly uniformly distributed across the output facet, $\mathrm{LG}_{\pm10}$ exhibit asymmetric power profiles, with opposite skew directions. Comparing the correlation matrices (side slices in the volumetric plot), stronger structural similarity is observed between $\mathrm{LG}_{00}$ and $\mathrm{LG}_{-10}$ for $\bigl\{(\upsilon, \nu) \mid \upsilon = \nu < 0\bigr\}$, and between $\mathrm{LG}_{00}$ and $\mathrm{LG}_{+10}$ for $\bigl\{(\upsilon, \nu) \mid \upsilon = \nu > 0\bigr\}$. These results demonstrate that while the power distribution across the fiber face varies, the inner product structure remains stable under deformation, providing insight into the fiber’s modal evolution and structural stability.%


\begin{figure*}[t!]%
  \subcaptionbox{\label{fig:approximatedLPmodes}}{\includegraphics[width=0.62576761\textwidth]{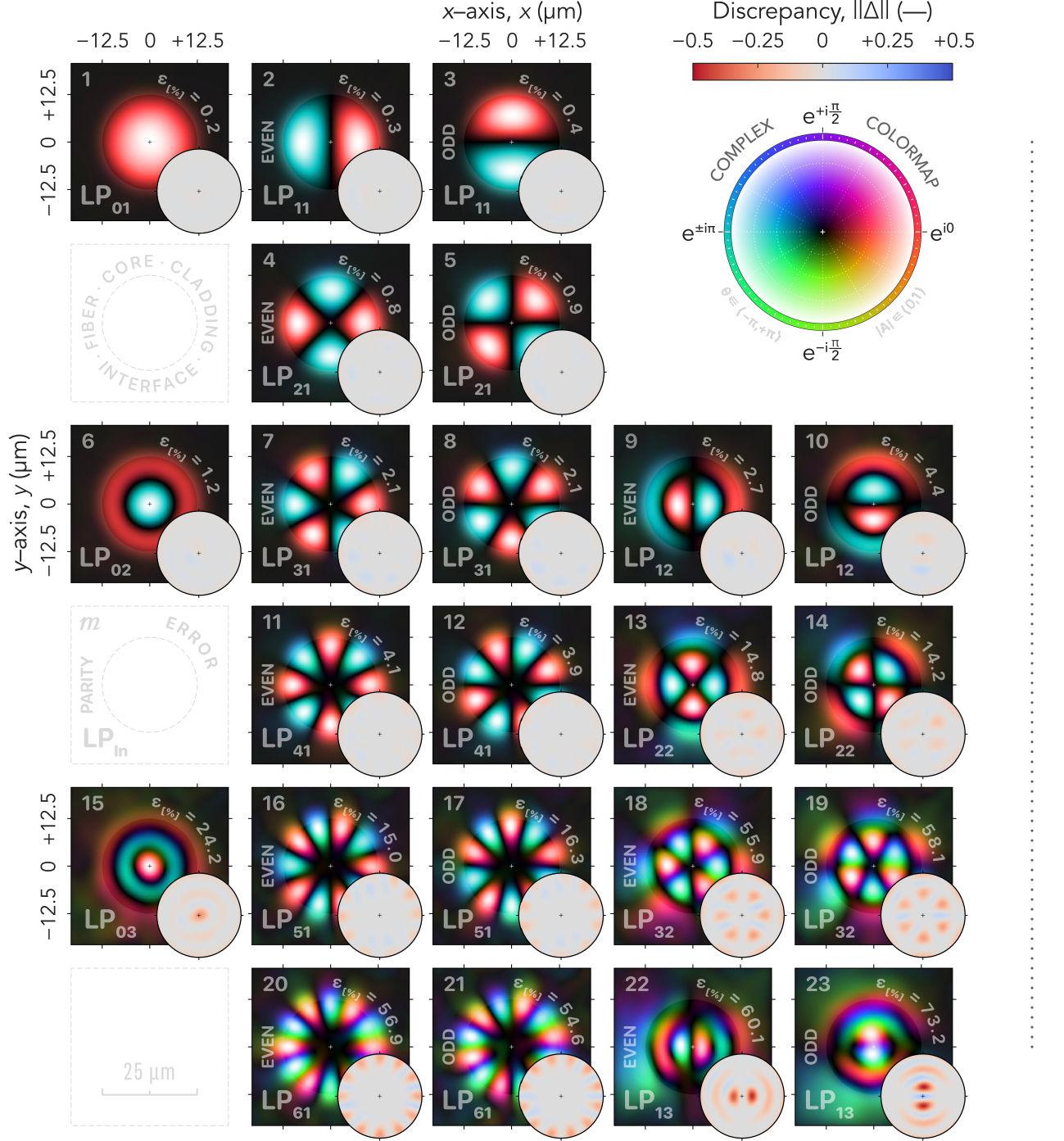}}\hfill%
  \subcaptionbox{\label{fig:matricesLPmodes}}{\includegraphics[width=0.3742349\textwidth]{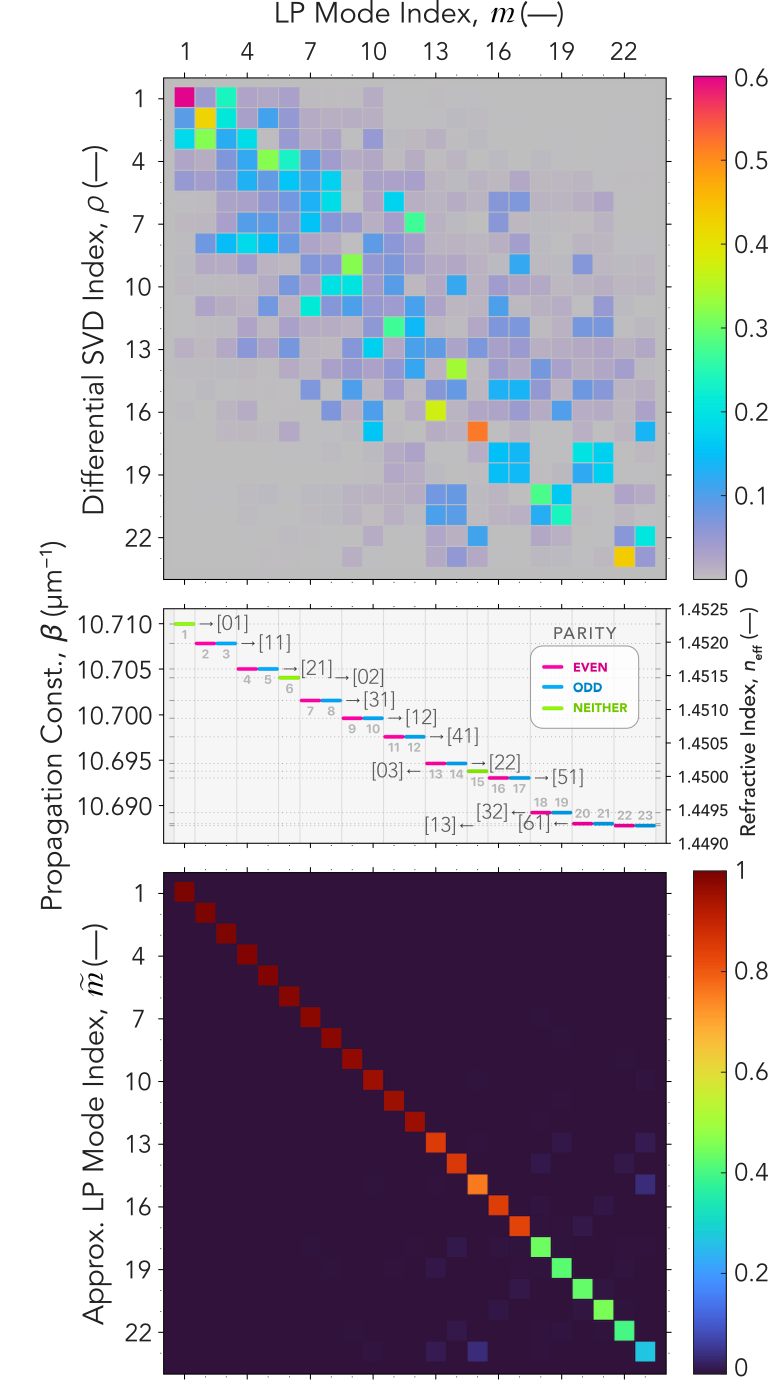}}%
	\caption{\label{fig:approximatedLPmodes_all}Projection of the theoretically calculated LP modes onto the bend-resolved singular basis ($\widehat{\Psi}_{\rho}$) inferred from experimental measurements. (a)~Spatial profiles of the LP modes, expressed as linear superpositions of the experimentally inferred bend-resolved modes, are shown alongside the normalized discrepancies relative to the ideal theoretical profiles. The visualization includes the complex colormap used throughout the manuscript, representing phase as hue and amplitude as a~linear transition from black to white. (b)~Top: Transformation matrix representing the correspondence between the theoretical LP modes and the experimentally determined bend-resolved common singular basis. Middle: Visualization of the modal propagation constants ($\beta_\mathcal{m}$) alongside their corresponding effective refractive indices ($n_\mathcal{m}^{\text{eff}}$). Bottom: Error matrix quantifying the discrepancies between the reconstructed LP basis and the ideal theoretical LP modes, displayed on a~linear scale and demonstrating pronounced diagonal dominance. Note:  $N_\mathrm{M} = \num{23}$ at $\lambda=\qty{852}{\nano\metre}$, ordered pairs of angular ($l$) and radial ($n$) mode numbers, where $l \in \mathbb{Z}$ and $n \in \mathbb{N}$, are mapped to a global index $\mathcal{m} \in \mathbb{N}$ via a~bijective function $\mathcal{m} = \mathcal{m}(l, n)$.}%
\end{figure*}%

\section{Alignment with Theoretical Eigensolutions}%

\noindent The theoretical number of guided LP modes was determined to be $N_\mathrm{M} = 23$ for light injection at $\lambda=\qty{852}{\nano\metre}$, based on the fiber’s parameters and tolerances specified by the manufacturer, \ch{SiO2} core with $\dmtr_{\text{core}}=\qty{25(3)}{\micro\metre}$, \ch{F:SiO2} cladding with $\dmtr_{\text{cladd}}=\qty{125(2)}{\micro\metre}$, $\mathrm{N\!A} = \num{0.100(0.015)}$. Using the manufacturer’s mean values, this result was obtained by analytically solving the eigenvalue equation under the weak guidance approximation and was found to be consistent with the experimentally extracted SVD modes, see~\cref{fig:approximatedLPmodes_all}. 


\section{\label{sec:experimental_implementation}Experimental Implementation}%


\noindent A~simplified schematic diagram of the experimental optical arrangement is shown in \cref{fig:optical_setup}. A~volume-holographic-grating (VHG) wavelength-stabilized laser diode with a~narrow linewidth operating at $\lambda=\qty{852(1)}{\nano\metre}$ was used as the light source. The beam was collimated by a~fixed-focal-point air-spaced doublet after passing through a~$\qty{1}{\metre}$ long SM polarization-maintaining fiber (PMF) terminated on both ends with FC/APC connectors. The subsequent half-wave plate (HWP) rotates the linearly polarized beam to be parallel with the extraordinary axis of the nematic liquid crystals on silicon (LCoS) of the SLM, achieving a~solely voltage-dependent phase shift with a~maximum diffraction efficiency. A~pair of achromatic positive doublet lenses (L\num{1}, L\num{2}) are used to expand the beam to sufficiently overfill the available $\qtyproduct{15.36x9.60}{\milli\metre}$ active area of the SLM chip with near-uniform illumination. After filtering the $\num{0}^{\text{th}}$ diffraction order in the Fourier domain with an aperture between the lenses L\num{3} and L\num{4}, the collimated beam of targeted $\num{1}\!^{\text{st}}$ diffraction order was demagnified with an another lens pair (L\num{5}, L\num{6}) to closely match the acceptance angle of the fiber core. The generated holograms positioned at the centre of the SLM’s Fourier plane, with a~resolution of $\qtyproduct{1900x1200}{\pixel}$ and a~pixel pitch of $\qtyproduct{8x8}{\micro\metre}$, are projected onto the proximal (near) facet of the fiber via an achromatic microscope objective. This objective has a~magnification of $\num{20}\times$, where the mode field diameter (MFD) of the fiber was digitally controlled via holograph spatial amplitude modulation to precisely adjust the diameter of the light beam coupled into the fiber. Complex spatial modulation allowed for fiber coupling of arbitrary base function, such as Laguerre--Gaussian ($\mathrm{LG_{\pm\ell\mathcal{p}}}$) or Hermite--Gaussian ($\mathrm{HG_{\mathcal{n}\mathcal{m}}}$). The fiber had a~total length of $L_\mathrm{T} = (\num[retain-zero-uncertainty=true]{1.000(0.075:0.000)})\,\unit{\metre}$ including the FC/PC connectors. Overfilling the numerical aperture (NA) can lead some incident light being coupled into cladding (i.e.~radiation) modes that can sufficiently propagate even over longer distances of fiber, leading to increased background in the output wavefront emitted at the distal (far) end of the fiber. The fiber is deliberately arranged in a~U-shaped configuration with permanently clamped center point to achieve predictable planar bends exclusively on its semicircular arc sections either side of this fixed point (i.e., the $y$-axis of the fiber remains in the $xz$-plane), as detailed further in~\ref{sec:bending_geometry}. A~motorised translation stage (MTS) with a~linear travel range $\Delta{p}_\text{mts}=\qtyrange[range-units=brackets,range-phrase=\text{--}]{0}{50}{\milli\metre}$ was used for controlled and highly repeatable bidirectional deformation (compression--expansion) of the fiber with an~accuracy of $\varepsilon_{\text{mts}}=\pm\qty{1.6}{\micro\metre}$. Subsequently, the speckled near-fields arising at the distal facet, with imprinted fiber conformations resulting from the stepwise movement of the MTS, are projected onto a~camera (CAM) through a~short focal length aspheric lens (L\num{7}) combined with a~long focal length achromatic doublet (L\num{8}). Before that, the object beam is combined with an external reference beam on a~non-polarizing beamsplitter (BS\num{2}) at a~mutually non-zero axial angle to enable fast holographic complex field acquisition from single-frame interferograms. The reference beam, taken from a~non-polarizing beamsplitter (BS\num{1}), is deliberately delayed by a~pair of mirrors to minimize the optical-path difference (OPD), considering the phase delay introduced by the fiber length. It is then expanded via a~pair of lenses (L\num{9}, L\num{10}) to achieve a~near-uniform planar wavefront for off-axis digital holography. Absorptive filters (ND\num{1}, ND\num{2}) are used accordingly to achieve high-dynamic mutual modulation while avoiding overexposure in the region of interest (ROI) within the camera’s full-frame resolution of $\qtyproduct{5328x3040}{\pixel}$. The incoherent light source (LED) with $\lambda_{0} = \qty{850}{\nano\metre}$ and bandwidth $\Delta\lambda = \qty{55}{\nano\metre}$ is used to minimize defocus in the projected wide-field image of the fiber’s distal facet on the CAM. As an extended source, it is only roughly collimated with an aspheric condenser lens (L\num{11}) and a~long focal length achromatic doublet (L\num{12}). The projection of the proximal facet onto the CAM is achieved in the same manner, even though it is not shown in the schematic for simplicity. Note that all lenses were experimentally aligned to ensure correct image replay, even though the focal length proportions are not accurately represented in~\cref{fig:optical_setup}.%


\begin{figure}[b!]%
	\subcaptionbox{\label{fig:fiberGeometry_shape}}{\includegraphics[width=0.8\columnwidth]{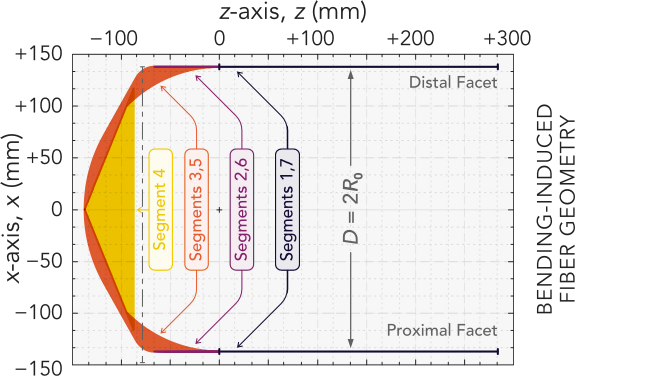}}\\%
	\subcaptionbox{\label{fig:fiberGeometry_plot}}{\includegraphics[width=0.8\columnwidth]{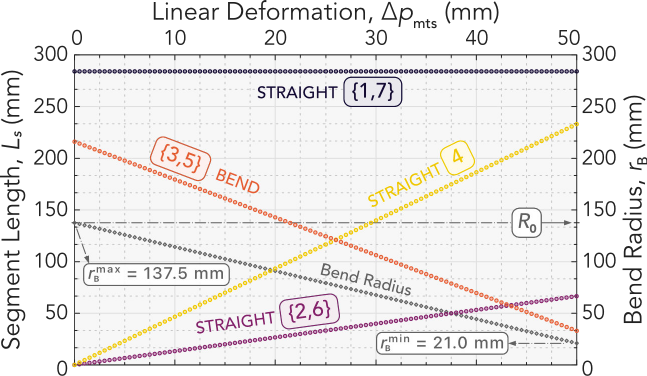}}%
	\vspace{-0.75em}%
	\caption{\label{fig:fiberGeometry}Deformation path of a~U-shaped fiber under controlled bending, constrained by~\cref{eq:bend_radii,eq:segment_lengths,eq:total_length}. (a)~Geometric layout of the U-shaped fiber in Cartesian coordinates, with segment lengths defined by the physical dimensions of the experimental setup. (b)~Plot of bend radius and segment length as functions of the applied deformation parameter $\Delta p_\mathrm{mts}$, set by the motorized translation stage. The bend radius applies only to bent segments $s = \{3, 4\}$, while $s = \{1, 2, 4, 6, 7\}$ are straight with effectively infinite radius.}%
\end{figure}%


\section{\label{sec:bending_geometry}Geometry of the Fiber Bending Path}%


As previously outlined, ensuring predictable deformation of the fiber during experiments, as demonstrated in \cref{fig:setup_bend,fig:fiberGeometry}, is pivotal for achieving precise and reproducible approximation of measurements. The fiber is permanently mounted in two parallel V-groove holders, equally separated by distance $D=2R_{0}$ in $xz$-plane, to achieve an ideal semicircular arc section of constant curvature $\rho(R_{0}) \propto R_{0}^{-1}$, defined by a~nominal radius $R_{0} = \qty{137.5}{\milli\metre}$. This constrains the fiber to a~specific geometric shape that can be longitudinally deformed on the flexible semicircular segment in a~deterministic manner between the inner edges of both holders acting as static nodes, eliminating the need for any additional movement of optical parts during its deformation (see~\cref{fig:fiberGeometry_shape}). The bidirectional deformation is conducted using a~linear MTS displacement ($\Delta{p}_\mathrm{mts}$) with a~one-sided V-groove restrainer holding the fiber at its midpoint during the compression--expansion process. This U-shaped configuration preferentially creates additional straight segments connected via two quarter-circle arcs with a~constant curvature, as illustrated in~\cref{fig:geometrical_bend,fig:fiberGeometry_shape}. The entire dynamic process can be encoded through a~constant number of segments $(N_\mathrm{seg} = \num{7})$ with precisely defined curvatures and lengths, where segments $s \in \{3, 5\}$ are bent, while segments $s \in \{1, 2, 4, 6, 7\}$ remain straight. Considering the initial semicircular geometry, the radius of the bent segments ($r_\mathrm{B}$) decreases linearly with the linear deformation applied by MTS displacement as follows
\begin{equation}\label{eq:bend_radii}
	r_{\mathrm{B}}\!\left(\Delta{p}_{\mathrm{mts}}\right) = R_{0} - \Delta{p}_{\mathrm{mts}}{\left(2-\frac{\uppi}{2}\right)}^{-1},
\end{equation}
until it gradually reaches the minimum $r_\mathrm{B}^\mathrm{min}=\qty{21.0}{\milli\metre}$, that is determined by the travel range limit of the MTS. Accordingly to radius decay given by \cref{eq:bend_radii}, the dynamic variation in the length of each segment  ($L_{s}$) triggered by $\Delta{p}_\mathrm{mts}$ can be mathematically expressed as
\begin{equation}\label{eq:segment_lengths}
L_{s}^{(p)}\!\left(r_{\mathrm{B}}\right) =
\begin{cases}
\:\dfrac{1}{2}(L_{\mathrm{T}} - \uppi R_{0}) & \text{if } s \in \{1, 7\}, \\[1.5ex]
\:\left(R_{0} - r_{\mathrm{B}}^{(p)}\right)\left(\dfrac{\uppi}{2} - 1\right) & \text{if } s \in \{2, 6\}, \\[1.5ex]
\:\dfrac{1}{2} \uppi r_{\mathrm{B}}^{(p)} & \text{if } s \in \{3, 5\}, \\[1.5ex]
\:2\left(R_{0} - r_{\mathrm{B}}^{(p)}\right) & \text{if } s = 4,
\end{cases}
\end{equation}%
while the cumulative sum of all segmental lengths must preserve the total length of the fiber for any arbitrary deformation of $\Delta{p}_\mathrm{mts}$, strictly following the expression
\begin{equation}\label{eq:total_length}
L_{\mathrm{T}} = \sum_{s=1}^{N_\mathrm{seg}} L_{s}^{(p)} \quad \forall\; p \in \{1, \dots, N_\mathrm{B}\}.
\end{equation}%
The bending limit in \cref{fig:geometrical_bend} can be achieved if the bent segments radius, centred and linearly shifting over the hypotenuse with a~defined length of $R_{0}\sqrt{1+(2-\uppi/2)^{2}}$, theoretically reaches $r_\mathrm{B}\!\bigl(R_{0}\left(2-\uppi/2\right)\bigr) = 0$. This implies that the curved segments disappear as $\rho(r_\mathrm{B})\rightarrow\infty$, theoretically bending the neutral axis of the fiber at a~right angle (discontinuous bend), resulting in the middle and side straight segment lengths of $L_{4}=2R_{0}$ and $L_{\{2,6\}} = R_{0}(\uppi/2-1)$. The semi-transparent area in \cref{fig:geometrical_bend} approximates the span covered in the $xz$-plane by the movement of the linear MTS, with curvatures in the range $\rho \approx ({7.\overline{27}}\,\text{–}\,{47.61})\,\unit{\per\metre}$, implying that the curvatures are much larger in magnitude than the core radius ($\rho\gg r_\text{core}$, where $r_\text{core} = \qty{12.5}{\micro\metre}$). The experiments were conducted at discrete positions $\Delta p_{\text{mts}} = \bigl\{\delta{p}_\mathrm{mts}\cdot(p-1)\;|\;p \in \mathbb{N},\, p \leq N_\mathrm{B}\bigr\}\,\unit{\milli\metre}$, where $\delta{p}_\mathrm{mts}=\qty{0.5}{\milli\metre}$, corresponding to the positions of the V-groove fiber restrainer in the given coordinate system $z = \bigl\{\Delta p_{\text{mts}} - R_{0} \bigr\}\,\unit{\milli\metre}$. The number of bends $N_\mathrm{B} = 101$ ensures a~thorough evaluation of the bending effects over the specified range, adequately capturing the propagation dynamics of the modes. This comprehensive linear stage translation enables precise tracking of the influence of each deformation position on the overall behavior of modal interactions within a~fiber.%


\section{\label{sec:transmission_throught_fiber}Transmission Through Fiber}%
Building on the previously established geometry based on segmenting the fiber into discrete sections with predictable lengths and curvatures, the dynamic alterations of an optical fiber bending path in a~U-shape configuration can be rigorously analyzed. By leveraging deterministic changes in lengths ($L_{s}$) and curvatures ($\rho_{s}$) into each segment modeled by its own TM, the resulting transmission matrix as a~function of altering $\Delta{p}_\text{mts}$ for the entire system can be dynamically described as their product
\begin{equation}\label{eq:total_transmission}
\mathbf{T}(\Delta{p}_\mathrm{mts}) = \prod_{s=N_\mathrm{seg}}^{1} \mathbf{T}_{s}^{(p)}(L_{s}, \rho_{s}), \quad \text{where} \; s \in \mathbb{N}.
\end{equation}
The dependence of the total TM, as given by~\cref{eq:total_transmission}, on the fiber’s segmental changes in length and curvature as a~function of the applied displacement by $\Delta{p}_\text{mts}$ can be fully expressed by \cref{eq:straight_transmission,eq:bend_transmission}. In the initial fiber configuration, with the largest radius of curvature, the TM is expressed as a~product of four contributing segments $\mat{T}(\Delta{p}_\text{mts}=0) = \mat{T}_{7}\mat{T}_{5}\mat{T}_{3}\mat{T}_{1}$, while the remaining segments $\mat{T}_{\left\{6,4,2\right\}}\equiv\mat{I}$, signify no contribution to the overall transmission. Considering that the fiber supports a~finite number of modes ($N_\mathrm{M}$) at a~certain wavelength, where each mode propagates over a~given length by its own specific modal propagation constant ($\beta_\mathcal{m}$), the TMs for straight and bend segments for each deformation of $\Delta{p}_\mathrm{mts}$ with a~dimension of $\mat{T}_{s}^{(p)} \in \mathbb{C}^{N_\mathrm{M} \times N_\mathrm{M}}$ can be, in idealized case, mathematically represented as
\begin{widetext}%
\begin{align}\label{eq:straight_transmission}
\underbracket[0.25pt][1.618pt]{\mathbf{T}_{s}^{(p)}(L_{s}, \rho_{s} = \infty)}_{\note{straight}} &=
\begin{bmatrix}
\begin{array}{*{4}{w{c}{3em}}}
\upe^{\upi\beta_{1} L_{s}^{(p)}} & 0 & \cdots & 0 \\
0 & \upe^{\upi\beta_{2} L_{s}^{(p)}} & \cdots & 0 \\
\vdots & \vdots & \ddots & \vdots \\
0 & 0 & \cdots & \upe^{\upi\beta_{N_{\mathrm{M}}} L_{s}^{(p)}}
\end{array}
\end{bmatrix} = \upe^{\upi\mat{B} L_{s}^{(p)}},%
\\[0.5em]
\underbracket[0.25pt][1.618pt]{\mathbf{T}_{s}^{(p)}(L_{s}, \rho_{s} \neq \infty)}_{\note{bend}}  &= 
\begin{bmatrix}\label{eq:bend_transmission}
\begin{array}{*{4}{w{c}{3em}}}
\tau_{11} & \tau_{12} & \cdots & \tau_{1N_{\mathrm{M}}} \\
\tau_{21} & \tau_{22} & \cdots & \tau_{2N_{\mathrm{M}}} \\
\vdots & \vdots & \ddots & \vdots \\
\tau_{N_{\mathrm{M}}1} & \tau_{N_{\mathrm{M}}2} & \cdots & \tau_{N_{\mathrm{M}}N_{\mathrm{M}}}
\end{array}
\end{bmatrix} = \upe^{\upi\widetilde{\mat{B}}L_{s}^{(p)}} \approx \sum_{\mathcal{m}=1}^{N_\mathrm{M}} \frac{1}{(\mathcal{m}-1)!} \left[\upi\widetilde{\mat{B}}L_s^{(p)}\right]^{\mathcal{m}-1},%
\end{align}
\end{widetext}%
where $\mat{B} = \text{diag}\left(\beta_{1}, \beta_{2}, \dots, \beta_{{N}_\mathrm{M}}\right)$ is a diagonal matrix of propagation constants applying phase shifts to each mode independently, while $\widetilde{\mat{B}}$ includes off-diagonal coupling terms, requiring a~full matrix exponential for mode evolution. Since the matrix $\widetilde{\mat{B}}$ is diagonalizable, its eigenvalues correspond to the effective propagation constants, while its eigenvectors represent the change in the modal basis in the perturbed fiber. In the first-order approximation, the elements of $\widetilde{\mat{B}}$ can be expressed as $\widetilde{\mat{B}} = \mat{B} + \mat{K}$, where $B_{\mathcal{m}\mathcal{m^{\prime}}} = \beta_\mathcal{m} \delta_{\mathcal{m}\mathcal{m^\prime}}$, representing the diagonal terms corresponding to the propagation constants of the uncoupled modes. The coupling matrix capturing bending-induced intermodal interactions is defined as $K_{\mathcal{m}\mathcal{m^{\prime}}}=\kappa_{\mathcal{m}\mathcal{m^{\prime}}}(1-\delta_{\mathcal{m}\mathcal{m^\prime}})$, where the off-diagonal coupling coefficients that interact nonlinearly in the exponential function are given by $\kappa_{\mathcal{m}\mathcal{m^{\prime}}} = -n_\text{core} k_0 \xi_{\mathrm{c}}\rho^{-1}_{s} \braketo{\Xi_\mathcal{m}}{\mathcal{x}}{\Xi_\mathcal{m^\prime}}$ for $\mathcal{m}\neq\mathcal{m^{\prime}}$. These off-diagonal coupling terms are inversely proportional to the curvature of the fiber and depend on the overlap integral between the $\mathcal{m}$-th and $\mathcal{m^\prime}$-th mode, weighted by the position coordinate $\mathcal{x}$ along the bend direction~\cite{Ploschner:2015}. Additionally, the correction factor $\xi_\mathrm{c}$ incorporates the photoelastic deformation tensor, accounting for lateral strains relative to longitudinal changes in fused silica, as determined by the Poisson ratio.%


\onecolumngrid

\newpage
\section{\label{sec:list_of_symbols}List of Symbols}%

\begin{table}[h!]%
\caption{\label{tab:list_of_symbols}List of symbols and physical quantities used throughout the manuscript, ordered alphabetically.}%
\renewcommand{\arraystretch}{1.15}%
\begin{minipage}[t!]{0.475\textwidth}%
\vspace{0pt}
\raggedright%
\footnotesize%
\begin{ruledtabular}%
\begin{tabular}{lll}%
\scriptsize\sffamily\bfseries{Symbol} &%
\scriptsize\sffamily\bfseries{Description} &%
\scriptsize\sffamily\bfseries{Unit}\vspace{2pt}\\%
\hline%
$\mathbf{B}$ & diagonal propagation constant matrix & $\unit{\per\metre}$ \\%
$\widetilde{\mathbf{B}}$ & perturbed matrix with intermodal coupling & $\unit{\per\metre}$ \\%
$\beta_\mathcal{m}$ & propagation constant of the $\mathcal{m}$-th fiber mode & $\unit{\per\metre}$ \\%
$\mat{C}_\mu$ & correlation matrix of speckles at fixed input $\mu$ & \--- \\%
$\chi_{\eta}^{(p)}$ & cumulative ratio up to $\eta$-th value at state $p$ & \--- \\%
$d$ & grating period of the SLM phase ramp & $\unit{\micro\metre}$ \\%
$D$ & distance between fiber V-groove holders & $\unit{\milli\metre}$ \\%
$\delta_{\mu\eta}$ & Kronecker delta, where $\delta_{\mu\eta} = [\mu = \eta]$ & --- \\%
$\delta{p}_\mathrm{mts}$ & uniform deformation step size  & $\unit{\milli\metre}$ \\
$\delta p_\mathrm{mts}^{[i]}$ & uniform deformation step size in sparse set $i$ & $\unit{\milli\metre}$ \\%
$\overline{\delta}_\mathrm{mts}$ & displacement accuracy of the MTS & $\unit{\micro\metre}$ \\%
$\Delta{p}_\mathrm{mts}$ & linear deformation applied by the MTS & $\unit{\milli\metre}$ \\%
$\Delta \widetilde{p}_\mathrm{mts}^{[i]}$ & sparse deformation set indexed by $i$ & $\unit{\milli\metre}$ \\%
$\Delta \breve{p}_\mathrm{mts}^{[i]}$ & subsampled deformation set index $i$ & $\unit{\milli\metre}$ \\
$\Delta\lambda$ & bandwidth of the incoherent LED source & $\unit{\nano\metre}$ \\%
$\dmtr_{\text{cladd}}$ & fiber cladding diameter & $\unit{\micro\metre}$ \\
$\dmtr_{\text{core}}$ & fiber core diameter & $\unit{\micro\metre}$ \\
$E$ & electric field (complex-valued) & --- \\%
$\epsilon$ & cutoff threshold in coupling strength & --- \\%
$\varepsilon_{\eta}^{(p)}$ & cumulative error up to $\eta$-th value at state $p$ & --- \\%
$\varepsilon_{[\unit{\percent}]}$ & per-mode reconstruction error (mode infidelity) & $\unit{\percent}$ \\%
$f_\mathrm{L}$ & effective focal length of preceding lens system & $\unit{\milli\metre}$ \\%
$\mathscr{F}^{[i]}$ & fidelity of sparse set or subset $i$ & $\unit{\percent}$ \\%
$\gamma_\mathrm{offset}^{[i]}$ & deformation offset for subset $i$ & $\unit{\milli\metre}$ \\%
$i$ & index of sparse set or subset & --- \\%
$k_0$ & free-space wavenumber & $\unit{\per\metre}$ \\%
$\mathbf{K}$ & off-diagonal modal coupling matrix & $\unit{\per\metre}$ \\%
$\mathcal{k}$ & index of singular value or corresponding vector & --- \\%
$\kappa_{\mathcal{m}\mathcal{m}'}$ & off-diagonal intermodal coupling coefficient & $\unit{\per\metre}$ \\
$\ell$ & azimuthal mode index & --- \\%
$L_\mathrm{T}$ & total length of the fiber & $\unit{\metre}$ \\%
$L_s^{(p)}$ & length of segment $s$ at deformation state $p$ & $\unit{\milli\metre}$ \\%
$\lambda$ & wavelength of the coupled laser & $\unit{\nano\metre}$ \\
$\lambda_0$ & central wavelength of incoherent LED source & $\unit{\nano\metre}$ \\
$\lambda_{\mathcal{k}}$ & $\mathcal{k}$-th eigenvalue of the diagonalized matrix & --- \\%
$\mat{\Lambda}$ & eigenvalue diagonal matrix & --- \\%
$\mathcal{m}$ & fiber mode index & --- \\%
$M_{\mu\eta}^{(p)}$ & entry indexed by $(\mu,\eta)$ in matrix at state $p$ & --- \\%
$\mat{M}^{(p)}$ & deformation-specific Hermitian correlation matrix & --- \\%
$\widehat{\mat{M}}$ & unified correlation matrix of singular modes & --- \\
$\mu, \eta$ & indices labeling modes or speckle patterns & --- \\%
$n$ & radial index of fiber mode & --- \\%
$n_\mathcal{m}^{\text{eff}}$ & effective refractive index of the $\mathcal{m}$-th mode & --- \\%
$n_\text{core}$ & refractive index of fiber core & ---\\%
$n_\mathrm{px}$ & number of pixels per $2\uppi$ phase ramp & --- \\%
$N_\mathrm{B}$ & number of bends or deformation states & --- \\%
$\breve{N}_\mathrm{B}$ & number of disjoint subset deformations & --- \\%
$\widetilde{N}_\mathrm{B}^{[i]}$ & number of sparse deformation states in set $i$ & --- \\%
$N_\mathrm{M}$ & number of guided modes in the fiber & --- \\%
$N_\mathrm{S}$ & number of speckle patterns involved & --- \\
$N_\mathrm{seg}$ & number of total fiber segments & --- \\
$p$ & deformation index related to $\Delta{p}_\mathrm{mts}$ & --- \\%
\end{tabular}
\end{ruledtabular}
\end{minipage}
\hfill
\begin{minipage}[t!]{0.475\textwidth}
\vspace{0pt}
\raggedright
\footnotesize
\begin{ruledtabular}
\begin{tabular}{lll}
\scriptsize\sffamily\bfseries{Symbol} &%
\scriptsize\sffamily\bfseries{Description} &%
\scriptsize\sffamily\bfseries{Unit}\vspace{2pt}\\%
\hline%
$p_\mathrm{SLM}$ & pixel pitch of the SLM & $\unit{\micro\metre}$ \\%
$\overline{P}$ & average output intensity (total) & --- \\%
$\overline{P}^{(p)}$ & average output intensity at state $p$ & --- \\%
$\Psi_{\eta}^{(p)}$ & singular mode at index $\eta$ for deformation $p$ & --- \\%
$\Psi_\varsigma$ & singular mode in concatenated set $\varsigma = \varsigma(\eta,p)$ & --- \\%
$\widehat{\Psi}_\rho$ & deformation-resolved global basis mode $\rho$ & --- \\%
$\breve{\Psi}_{\rho}^{[i]}$ & disjoint subset-derived singular mode in set $i$ & --- \\%
$\widetilde{\Psi}_{\rho}^{[i]}$ & sparse deformation-sampled basis mode in set $i$ & --- \\%
$\mathcal{r}$ & full rank of matrix & --- \\%
$\mathcal{r}_\mathrm{X}$ & effective modal rank (after thresholding) & --- \\%
$\mathcal{r}_{\widetilde{\mathrm{X}}}$ & rank estimate affected by noise or artifacts & --- \\%
$R_{0}$ & nominal bend radius in undeformed state & $\unit{\milli\metre}$ \\%
$\rho$ & index labeling global singular modes & --- \\%
$\rho_s$ &  curvature of fiber segment $s$ & $\unit{\per\metre}$ \\%
$r,\theta,z$ & polar coordinates: radius, angle, axial distance & $\unit{\micro\metre}$, $\unit{\radian}$, $\unit{\milli\metre}$\\%
$r_\mathrm{B}$ & radius of bent segments & $\unit{\milli\metre}$ \\%
$r_\mathrm{B}^\mathrm{min}$ & minimum bend radius & $\unit{\milli\metre}$ \\%
$r_\mathrm{B}^\mathrm{max}$ & maximum bend radius, equal to $R_{0}$ & $\unit{\milli\metre}$ \\%
$r_\text{core}$ & core radius of the fiber & $\unit{\micro\metre}$ \\%
$s$ & index of the fiber segment & --- \\%
$S_{\mu}^{(p)}$ & speckle field at input index $\mu$ and state $p$ & --- \\%
$\widehat{S}$ & normalized speckle field & --- \\%
$\sigma_{\mathcal{k}}^{(p)}$, $\sigma_{\eta}^{(p)}$ & singular value at state $p$, indexed by $\mathcal{k}$ or $\eta$ & --- \\%
$\widetilde{\sigma}_{\eta}^{(p)}$ & relative coupling strength at state $p$ & $\unit{\decibel}$ \\%
$\varsigma$ & global singular mode index, where $\varsigma = \varsigma(\eta,p)$ & --- \\%
$\mat{\Sigma}^{(p)}$ & diagonal matrix of singular values at state $p$ & --- \\%
$\mat{T}$ & total transmission matrix for the fiber & --- \\%
$\mat{T}_s^{(p)}$ & transmission matrix of segment $s$ at state $p$ & --- \\%
$\tau_{ij}$ & transmission matrix element (bend) & --- \\%
$\theta_\mathrm{c}$ & critical angle for total internal reflection & $\unit{\radian}$ \\
$V_{\mu\eta}^{(p)}$ & projection coefficient, input $\mu$, mode $\eta$, state $p$ & --- \\
$\mat{U}^{(p)}, \mat{V}^{(p)}$ & left/right singular matrices at state $p$ & --- \\
$\vek{u}_{\mathcal{k}}^{(p)}, \vek{v}_{\mathcal{k}}^{(p)}$ & left/right singular vectors at state $p$ & --- \\%
$\upsilon, \nu$ & indices of laterally displaced input modes & --- \\%
$x, y, z$ & Cartesian coordinate axes & $\unit{\milli\metre}$ \\%
$x_0, y_0$ & Cartesian coordinates at fiber center & $\unit{\micro\metre}$ \\%
$x_{\upsilon\nu}, y_{\upsilon\nu}$ & center coordinates of displaced input mode & $\unit{\micro\metre}$ \\%
$\xi_\mathrm{c}$ & photoelastic correction factor & --- \\%
$\Xi_\mathcal{m}$ &  theoretical eigenmode field of fiber mode $\mathcal{m}$ & --- \\%
$\widehat{z}$ & scaled axial propagation distance & $\unit{\milli\metre}$ \\
$\zeta_{\eta}^{(p)}$ & cumulative sum up to the $\eta$-th value at state $p$ & --- \vspace{10pt}\\
$|\cdot|$ & field amplitude (modulus) & --- \\
$|\cdot|^{2}$ & field intensity (modulus squared) & --- \\
$(\cdot)^{\ast}$ & complex conjugate & --- \\
$(\cdot)^\mathsf{H}$ & Hermitian (conjugate) transpose & --- \\
$\{\cdot\}$ & set notation & --- \\
$(\cdot,\cdot)$ & ordered pair & --- \\
$(\cdot,\cdot,\cdot)$ & ordered triplet & --- \\
$\big\{(\cdot,\cdot),(\cdot,\cdot)\big\}$ & set of ordered pairs & --- \\
$\|\cdot\|$ & Euclidean or general vector norm & --- \\
$\braket{\cdot}{\cdot}$ & inner product (Dirac bra–ket notation) & --- \\
\end{tabular}
\end{ruledtabular}
\vspace{5pt}
\end{minipage}
\end{table}
\vfill

\end{document}